\documentclass[12pt,a4paper]{article}
\usepackage{type1cm,amsmath,hangcaption,graphicx,indentfirst}
\usepackage[psamsfonts]{amssymb}
\numberwithin{equation}{section}

\newcommand{\br}{\mathbb R}

\DeclareMathOperator*{\arccosh}{{\rm arccosh}}
\DeclareMathOperator*{\arcsinh}{{\rm arcsinh}}

\begin{document}
%%% Title page %%%%%
\begin{titlepage}

 \renewcommand{\thefootnote}{\fnsymbol{footnote}}
\begin{flushright}
 \begin{tabular}{l}
 SNUST-050301\\
 ROM2F/2005/04 \\
 hep-th/0503148\\
 \end{tabular}
\end{flushright}

 \vfill
 \begin{center}
 \font\titlerm=cmr10 scaled\magstep4
 \font\titlei=cmmi10 scaled\magstep4
 \font\titleis=cmmi7 scaled\magstep4
 \centerline{\titlerm D-branes in a Big Bang/Big Crunch Universe:}
 \vskip .3 truecm
 \centerline{\titlerm Nappi-Witten Gauged WZW Model}
 \vskip 2.5 truecm

\noindent{ \large Yasuaki Hikida,$^a$\footnote{E-mail:
hikida@phya.snu.ac.kr} Rashmi R.~Nayak$^b$\footnote{E-mail:
Rashmi.Nayak@roma2.infn.it} and Kamal
L.~Panigrahi$^b$\footnote{E-mail: Kamal.Panigrahi@roma2.infn.it}}
\bigskip

 \vskip .6 truecm
\centerline{\it $^a$ School of Physics \& BK-21 Physics Division,
                Seoul National University}
\centerline{\it Seoul 151-747, Korea}
\bigskip
\centerline{\it $^b$ Dipartimento di Fisica \& INFN, Sezione di
Roma 2, ``Tor Vergata'',} \centerline{\it Roma  00133, Italy}

 \vskip .4 truecm

 \end{center}

 \vfill
\vskip 0.5 truecm

\begin{abstract}

We study D-branes in the Nappi-Witten model, which is a gauged WZW
model based on $(SL(2,\br) \times SU(2)) / (U(1) \times U(1))$.
The model describes a four dimensional space-time consisting of
cosmological regions with big bang/big crunch singularities and
static regions with closed time-like curves. The aim of this paper
is to investigate by D-brane probes whether there are pathologies
associated with the cosmological singularities and the closed
time-like curves. We first classify D-branes in a group theoretical way,
and then examine DBI actions for effective theories on the D-branes. In
particular, we show that D-brane metric from the DBI action does
not include singularities, and wave functions on the D-branes are
well behaved even in the presence of closed time-like curves.

\end{abstract}
\vfill
\vskip 0.5 truecm

\setcounter{footnote}{0}
\renewcommand{\thefootnote}{\arabic{footnote}}
\end{titlepage}

\newpage

\tableofcontents
%%%%%%%%%%%%%%%%%%%%%%%%%%%%%%%%%%%%%%%%%%%%%%%%%%%%%%%%%%%%%%%%%%%%%%
\newpage
\section{Introduction}
\label{Intoduction}

The formulation of string theory in cosmological or broadly
time-dependent backgrounds still remains as one of the most
exciting open challenges for theoretical physicists. Among other
things, it is important to see how (whether) string theory resolves the
cosmological singularities (or not). Some recent attempts along this
direction have been made in
\cite{KOSST,Seiberg,BHKN,CC,Nekrasov,Simon,LMS1,CCK,Lawrence,LMS2,FM,HP,BCKR}.
In this paper, we investigate the Nappi-Witten (NW)
model \cite{NW,NW2} as a conformal field theory (CFT) model
that includes big bang and big crunch singularities. This model is defined
as an $SL(2,\br) \times SU(2)$ WZW model gauged by $U(1) \times
U(1)$ in an asymmetric way. Therefore we can solve it exactly and gain
some understanding of the behaviour of strings across these singularities.
Similar attempts have been made to understand the cosmological backgrounds
by using gauged WZW models, e.g., in \cite{Horava,KL,CKR,HikiTaka,TT}.

The NW model describes a four dimensional anisotropic
universe that starts at a big bang singularity, expands for a
while, ends at a big crunch singularity within a finite affine
time, and again starts at another big bang singularity.
It was argued in \cite{NW2} that strings across these
singularities make sense because the theory is defined
as a coset CFT. This may be related to the fact that strings
have a finite length scale $l_s$ and may go through the
singularities. As described in \cite{NW2},
the coset contains, in addition to copies of the cosmological regions,
non-compact and time independent (static) regions known as {\it whiskers}.
The whisker part of the full
space-time contains closed time-like curves (CTCs), which
seems to be unacceptable for any measurable quantity like S-matrix.
Nevertheless, these CTCs are argued to be necessary for defining the
observables in terms of gauged WZW model \cite{NW2}.

Hence we would like to ask a question,
whether studying the dynamics of D-branes sheds more lights on the
understanding of big bang/big crunch singularities and CTCs.
This is a non-trivial question because the D-branes, or the open strings ending on
them, feel the background geometry in a different way.
In particular, D0-brane is an useful candidate to probe the
singularities, because the point particle could probe directly the
singular points. In fact, the singularities are at strong coupling
regions, thus D-branes may serve as more effective probes than strings.
Moreover, we might expect that the effective theories on D-branes
wraping the CTCs include pathologies associated with CTCs.
This is an analogous situation to D-branes in a supersymmetric
G\"{o}del-type universe \cite{DFS1}.%
\footnote{See, e.g., \cite{GGHPR,BGHV,HT,DFS1,HR,BDPR,BHH,JR}
for recent discussions on CTCs in supersymmetric
G\"{o}del-type universes.}

The organization of the paper is as follows. In section \ref{BG},
we review the geometry of the quotient CFT background $(SL(2,
{\mathbb R})\times SU(2))/(U(1) \times U(1))$ \cite{NW,NW2},
where we fix the notation and summarize the properties of the geometry.
In section
\ref{D-branec}, we investigate D-branes in the cosmological
regions. Branes in the coset theory are constructed to descend
from maximally symmetric and symmetry breaking branes
in $SL(2,\br) \times SU(2)$ WZW models \cite{Sarkissian,QS}.
We use the DBI analysis to investigate by D-brane probes
the geometry especially near the singularities.
In particular, we show that the singularities are cancelled
in the DBI action, so the D-brane metric does not include
singularities.
We also find that the classical solutions reproduce the
group theoretic results.
In section \ref{D-branew}, we investigate D-branes in the whisker
regions. The trajectories of the D-branes are also classified in a
group theoretical way, and the DBI action is used to probe the
geometry by the branes. Open string spectrum on the D-brane wrapping CTCs can be
read off from the eigenfunction of Laplacian in terms of open
string metric and open string coupling \cite{SW}. We see that
the spectrum does not include any pathology associated with CTCs.
Section \ref{WF} is devoted to a detailed study of the wave functions
in various sectors of the model.
Finally, we
conclude in section \ref{conclusion} with discussions and some open
questions. Appendix \ref{Misner} includes a brief analysis on
D-branes in Misner space-time.

%%%%%%%%%%%%%%%%%%%%%%%%%%%%%%%%%%%%%%%%%%%%%%%%%%%%%%%%%%%%%%%%%%%%%%

\section{Background geometry}
\label{BG}

In general it is difficult to analyze string theory in
cosmological backgrounds. However, the gauged WZW models are few techniques to
construct exact string models for cosmological systems
(see, e.g., \cite{Horava,KL,NW,NW2,CKR,HikiTaka,TT}).
In this paper, we investigate a coset model proposed by
Nappi and Witten \cite{NW}, which describes a four dimensional anisotropic,
expanding and contracting closed universe.
The coset that we are interested in is $SL(2,\br) \times SU(2)$ WZW model
with the following identification\footnote{We only use a specific
identification adopted in \cite{NW2} among the general ones given in
\cite{NW}.}
\begin{align}
  (g,g') \sim
  (e^{\rho \sigma_3} g e^{\tau \sigma_3},
   e^{i \tau \sigma_3} g ' e^{i \rho \sigma_3}) ,
   \label{gauge}
\end{align}
where we denote $g,g'$ as elements of $SL(2,\br)$ and $SU(2)$,
respectively. The levels $k,k'$ for the $SL(2,\br)$ and $SU(2)$
WZW models are set to be equal $(k=k')$, and they are considered
to be large because we are interested in the classical
properties.\footnote{Hereafter we will not put explicitly $k$ and
$k'$'s in order to make various formulae simpler, though they can
be inserted back when needed.} A critical bosonic string theory
should include extra space dimensions such that the total central
charge $c = 26$. The extension to the superstring case should be
straightforward.

\begin{figure}
\centerline{\scalebox{.3}{\includegraphics{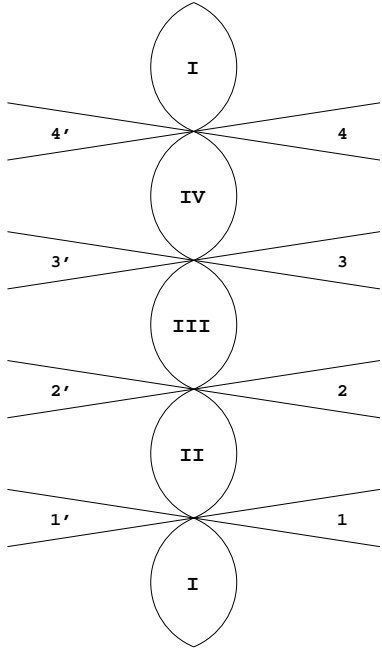}}} \caption{\it
A two-dimensional sketch of the four-dimensional space-time.
Regions $I-IV$ represent closed universe regions with the
singularities at the beginning and at the end. Regions $1-4$ and
$1'-4'$ are the whisker regions which contain closed time-like
curves.} \label{NW-universe}
\end{figure}
As we will see below, the NW model includes twelve regions as in
fig. \ref{NW-universe}. It originates from the fact that our
parametrization of $SL(2,\br)$ is divided into twelve cases,
depending on the signs of its group elements. We find it
convenient to represent, inside each regions, the most general
group element $g \in SL(2, \br)$ as
\begin{align}
  g &= e^{\alpha \sigma_3} (-1)^{\epsilon_1}
        (i \sigma_2)^{\epsilon_2} g_{\delta} (\theta)
     e^{\beta \sigma_3} ,
\end{align}
where $\alpha, \beta \in \br$ and $\epsilon_{1},\epsilon_{2}=0,1$.
If we set $\epsilon_{1}= \epsilon_{2} = 0$, then the parameters are
in region $\delta=II,1,1'$ with
\begin{align}
 g_{II} &=
    \begin{pmatrix}
     \cos \theta & \sin \theta \\
     - \sin \theta & \cos \theta
    \end{pmatrix} ,
 &g_{1} &=g_{1'}^{-1}=
    \begin{pmatrix}
     \cosh \theta & \sinh \theta \\
     \sinh \theta & \cosh \theta
    \end{pmatrix} .
\end{align}
The parameter $\theta$ runs from $0 ~{\rm to}~ \pi/2 ~(0 \leq \theta \leq \pi/2)$ in region II and
$\theta \geq 0$ in regions $1$ and $1'$. We obtain, in total,
twelve regions by changing $\epsilon_{1,2}=0,1$, however it is evident that the
change of $\epsilon_{1,2}$ does not affect the value of the quantity
\begin{align}
 W = {\rm Tr} (\sigma_3 g \sigma_3 g^{-1}).
\end{align}
\begin{table}
\begin{center}
\begin{tabular}{|c|c|cccc|}
\hline
 (A) & $W > 2$ &
  $1 \begin{pmatrix} \ + & + \\ + & + \end{pmatrix}$ &
  $1' \begin{pmatrix} \ + & - \\ - & + \end{pmatrix}$ &
  $3 \begin{pmatrix} \ - & - \\ - & - \end{pmatrix}$ &
  $3' \begin{pmatrix} \ - & + \\ + & - \end{pmatrix}$ \\ \hline
 (B) & $-2 < W < 2$ &
  $I \begin{pmatrix} \ + & - \\ + & + \end{pmatrix}$ &
  $II \begin{pmatrix} \ + & + \\ - & + \end{pmatrix}$ &
  $III \begin{pmatrix} \ - & + \\ - & - \end{pmatrix}$ &
  $IV \begin{pmatrix} \ - & - \\ + & - \end{pmatrix}$ \\ \hline
 (C) & $W < - 2$ &
  $2 \begin{pmatrix} \ - & + \\ - & + \end{pmatrix}$ &
  $2' \begin{pmatrix} \ + & + \\ - & - \end{pmatrix}$ &
  $4 \begin{pmatrix} \ + & - \\ + & - \end{pmatrix}$ &
  $4' \begin{pmatrix} \ - & - \\ + & + \end{pmatrix}$ \\
 \hline
\end{tabular}
\end{center}
\caption{\it Twelve regions for the parametrization of $g \in SL(2,\br)$.}
\label{12}
\end{table}
In other words, we can separate the twelve regions into three types
as in table \ref{12}.\footnote{We use the label of regions
adopted in \cite{NW2}.}
In type (B), the parameter
$\theta$ in regions $III, IV,$ and $I$ may be obtained by shifting
$\theta \to \theta + n \pi /2$ with $n = 1,2,3$. For $A$ and $A'$
with $A=1,2,3,4$, we may replace $\theta$ by $-\theta$. Moreover,
regions 1 and 3, or regions 2 and 4 are different only
with the overall factor $g \to -g$.

Similarly, for $g' \in SU(2)$ we can use for all the regions
\begin{align}
  g' = e^{i \beta ' \sigma_3}
        e^{i \phi \sigma_2}
        e^{i \alpha ' \sigma_3}
\end{align}
with $\alpha ' \sim \alpha ' + 2 \pi, \beta ' \sim \beta '+  2\pi$ and
$0 \leq \phi \leq \pi /2$.
Even so, we adopt a bit different way of
parametrization in order to express the metric in a simpler form as
we see below.
Moving into the regions $III,IV,$ and $I$,
 we change $\phi$ into $\phi \to \phi + n \pi/2$  $(n=1,2,3)$
and shift the region into $ -n\pi/2 < \phi < (1-n) \pi /2$. The
same thing should happen when we change $1 \to 2 \to 3 \to 4$ or
$1' \to 2' \to 3' \to 4'$. {}From now on, we will concentrate on
region $II$, region $1$ and region $2'$, since the other regions
can be analyzed by using the above transformations.

The geometry of the NW model can be obtained by
starting from the $SL(2,\br) \times SU(2)$ WZW action and then by integrating out
the gauge fields associated with $U(1) \times U(1)$ \cite{NW,NW2}.
The coordinates invariant under the gauge transformation \eqref{gauge}
are given by $\alpha - \alpha'$, $\beta - \beta '$, $\theta$, $\phi$.
We use a gauge choice $\alpha ' = \beta ' = 0$
and denote $\alpha \pm \beta = \lambda_{\pm}$.
In region $II$, metric, B-field and dilaton are given by%
\footnote{Exact metric, B-field and dilaton including all $1/k$
corrections can be computed by following \cite{BS}.
See also \cite{JS}.}
\begin{align}
&ds^2 = -d \theta ^2 + d {\phi}^2
  + \frac{\cot ^2 \phi}{ 1 + \tan ^2 \theta \cot ^2 \phi}
    d\lambda ^2 _+
  + \frac{\tan ^2 \theta}{ 1 + \tan ^2 \theta \cot ^2 \phi}
    d\lambda ^2 _- , \nonumber \\
&B_{\lambda_-  \lambda_+} = \frac{1}{1+\tan ^2 \theta \cot ^2 \phi} ,
 \qquad
\Phi = \Phi_0 - \frac{1}{2}
 \log (\cos ^2 \theta \sin ^2 \phi + \sin ^2 \theta \cos ^2 \phi).
 \label{metric}
\end{align}
The time coordinate is given by $\theta$, which starts at $\theta =
0$ and ends at $\theta = \pi /2$. At the beginning $(\theta = 0)$, a
component of the metric $g_{\lambda_- \lambda_-}$ vanishes,
therefore the volume of the universe also vanishes. We interpret this as
the universe starts at $\theta = 0$ with a big bang singularity.
At the end $(\theta = \pi /2)$, another component of the metric $g_{\lambda_+
\lambda_+}$ becomes zero, so this point may be interpreted as a
big crunch singularity. For a general $\phi$, these singularities are of
orbifold-type. However, at $\theta=\phi=0$ or $\theta=\phi=\pi/2$,
the metric and dilaton field diverge, and these singularities
are of the same type as the space-like singularities in the two dimensional
black hole \cite{witten,DVV}.
More detailed structure of these singularities in this
region can be found in \cite{NW}. The ranges of other parameters are
$0 \leq \phi \leq \pi /2 $ and $0 \leq \lambda_{\pm} < 2\pi$.
Note that $\lambda_{\pm}$ are periodic parameters
$(\lambda_{\pm} \sim \lambda_{\pm} + 2 \pi)$ due to the
periodicity of the parameters $\alpha ' ~{\rm and} ~\beta '$.
The constant part of the dilaton field will be set to be zero $(\Phi_0 = 0)$ for
simplicity.

As we will see later, wave functions are given as eigenfunctions of
Laplacian, which is given by
\begin{align}
\Delta = \frac{1}{e^{-2 \Phi} \sqrt{-g}}
   \partial_{\mu} e^{-2 \Phi} \sqrt{-g} g^{\mu \nu}
   \partial_{\nu}
\end{align}
in a general curved background.
The Laplacian in this region can be computed as
\begin{align}
 \Delta =& \left[
  - \frac{1}{\cos \theta \sin \theta} \partial_{\theta}
   (\cos \theta \sin \theta) \partial_{\theta} +
   \tan ^2 \theta \partial_{\lambda_+}^2 + \cot ^2 \theta
    \partial_{\lambda_-}^2
 \right] \nonumber\\ &+\left[
   \frac{1}{\cos {\phi} \sin {\phi}} \partial_{{\phi}}
   (\cos {\phi} \sin {\phi}) \partial_{{\phi}} +
   \tan ^2 {\phi} \partial_{\lambda_+}^2 + \cot ^2 {\phi}
    \partial_{\lambda_-}^2
 \right] ,
 \label{LBII}
\end{align}
which can be seen as a sum of two contributions coming from
$SL(2,\br)$ and $SU(2)$ parts.
Hence we can examine the wave functions for $SL(2,\br)$ and $SU(2)$ parts
separately.
In particular, we do not expect any peculiar behavior of wave functions
at $\theta=\phi=0$ or $\theta=\phi=\pi/2$.

In region 1, metric, B-field and dilaton are given by
\begin{align}
&ds^2 = d \theta ^2 + d {\phi}^2
  + \frac{\cot ^2 \phi}{ 1 - \tanh ^2 \theta \cot ^2 \phi}
    d\lambda ^2 _+
  - \frac{\tanh ^2 \theta}{ 1 - \tanh ^2 \theta \cot ^2 \phi}
    d\lambda ^2 _- ,\nonumber \\
&B_{\lambda_+  \lambda_-} =
 \frac{1}{1 - \tanh ^2 \theta \cot ^2 \phi} , \qquad
\Phi = - \frac{1}{4}
 \log (\cosh ^2 \theta \sin ^2 \phi
       -  \sinh ^2 \theta \cos ^2 \phi)^2 \label{metric1}.
\end{align}
The ranges of parameters are $0 \leq \theta$, $0 \leq \phi \leq \pi
/2 $ and $0 \leq \lambda_{\pm} < 2\pi$.
Because of the periodicity condition on $\lambda_{\pm}$, there are
always CTCs in this region.
It is important to note that there is a singular surface along the line
\begin{align}
 1 = \tanh ^2 \theta \cot ^2 \phi ,
 \label{singular-surface}
\end{align}
which may be seen as a domain wall, and across this line the time
coordinate is exchanged between $\lambda_+$ and $\lambda_-$.%
\footnote{The structure of this singular surface is
examined in detail in \cite{NW2}.}
Despite the complexity of the structure of this region,
the Laplacian is given in a rather simple form as
\begin{align}
  \Delta =& \left[
   \frac{1}{\cosh \theta \sinh \theta} \partial_{\theta}
   (\cosh \theta \sinh \theta) \partial_{\theta}
  - \tanh ^2 \theta \partial_{\lambda_+}^2 - \coth ^2 \theta
    \partial_{\lambda_-}^2
 \right] \nonumber\\ &+\left[
   \frac{1}{\cos {\phi} \sin {\phi}} \partial_{{\phi}}
   (\cos {\phi} \sin {\phi}) \partial_{{\phi}} +
   \tan ^2 {\phi} \partial_{\lambda_+}^2 + \cot ^2 {\phi}
    \partial_{\lambda_-}^2
 \right] .
 \label{LB1}
\end{align}
In particular, the singular surface \eqref{singular-surface} cannot be seen in this expression.
In region $2'$, we use the following parametrization
\begin{align}
  g &= e^{\alpha \sigma_3}
    \begin{pmatrix}
     \sinh \theta & \cosh \theta \\
     - \cosh \theta & - \sinh \theta
    \end{pmatrix}
     e^{\beta \sigma_3} .
\end{align}
Due to the change of $\phi$, metric, $B$-field and
dilation are given by just replacing $\lambda_+$ by $\lambda_-$ in
those for region 1.
We should also remember that $\phi$ runs from $-\pi/2~ {\rm to}~ 0~(-\pi/2 \leq \phi < 0)$.

\section{D-branes in the cosmological regions}
\label{D-branec}

Let us move to D-branes in the NW model and start from a cosmological
region (region $II$). In this region, the background is time-dependent,
and hence the energy is not conserved. D-branes in the coset theory are
constructed to descend from maximally symmetric and symmetry breaking
D-branes in $SL(2,\br) \times SU(2)$ WZW model, which are classified by
(twisted) conjugacy classes \cite{Sarkissian,QS}.
DBI actions are used to investigate
the background geometry, especially near the singularities.
We show that the classical solutions reproduce the group theoretical
results, and we also examine the wave functions on the D-branes by making use of
open string metrics \cite{SW}.

\subsection{Geometry of D-branes: A group theoretic view}
\label{conjugacyc}

Maximally symmetric and symmetry breaking D-branes in WZW models
are classified by conjugacy classes and by twisted conjugacy classes,
respectively \cite{KO,AS,BFS,BPPZ,FFFS}.
For these D-branes, holomorphic and anti-holomorphic currents
for bulk symmetry are related by linear equations at the
worldsheet boundary.\footnote{With this type of boundary conditions,
we can show that the conformal symmetry at the boundary is not broken.
There might be other types of D-branes which preserve the boundary
conformal symmetry, but we do not know how to deal with.
The situation for D-branes in cosets is similar.}
Therefore, the corresponding D-branes should
be invariant under the symmetry left,
which are characterized by (twisted) conjugacy classes.
D-branes in coset models can be constructed to descend from these
D-branes in a systematic way, for example, as in
\cite{MMS,Gawedzki,ES,FS,Ishikawa}, and for
asymmetric cosets including the NW model as in \cite{WZ,Sarkissian,QS}.
We obtain the geometry of D-branes using these methods as follows.

We first focus on the conjugacy classes and the twisted conjugacy classes of
$SL(2,\br)$ and of $SU(2)$. For $SL(2,\br)$ part, they are given by
\begin{align}
 &{\rm A}:\quad
  {\rm Tr}\, g = 2 \cosh (\alpha + \beta) \cos \theta = 2 \kappa ,\\
 &{\rm B}:\quad
 {\rm Tr}\, \sigma_1 g = 2 \sinh (\alpha - \beta) \sin \theta = 2 \kappa ,
\end{align}
where $\kappa \in \br$ and we denote them as A-type and B-type,
respectively. The $SL(2,\br)$ WZW model describes strings in $AdS_3$,
and the worldvolumes for A-branes are $H_2$ for $0 < \kappa < 1$,
light-cone for $\kappa = 1$ and $dS_2$ for $\kappa > 1$, and that
for B-brane is $AdS_2$ \cite{BP}. Similarly,
for $SU(2)$ part, the (twisted) conjugacy class is
given by
\begin{align}
 &{\rm A}:\quad
 {\rm Tr}\, g' = 2 \cos (\alpha ' + \beta ') \cos \phi = 2 \xi ,
  \label{SU2A}\\
 &{\rm B}:\quad
 {\rm Tr}\, \sigma_1 g'= - 2 i \sin (\alpha ' - \beta ') \sin \phi=
   - 2 i \xi , \label{SU2B}
\end{align}
where $\xi \in \br$ and we call them as A-type and B-type.
The twist in the B-type is just an inner automorphism of $SU(2)$
in this case, and hence, in the $SU(2)$ WZW model,
the B-brane is obtained by rotating the A-brane.
However, in the parafermionic theory $SU(2)/U(1)$, it has been shown that
gauging of $U(1)$ part leads to two inequivalent types of D-branes
\cite{MMS,Ishikawa}.

In order to construct D-branes in the gauged WZW model, we have to
take into account of the gauge symmetry. For the purpose,
we sum up all the gauge orbits of the (twisted)
conjugacy classes, project them into the gauge invariant space, and
finally shift by an element of $U(1) \times U(1)$. This procedure
is essentially the same as taking the product of conjugacy
classes of $SL(2,\br) \times SU(2)$ and $U(1) \times U(1)$ as
in \cite{QS}. Let us call them as XY-type D-brane or XY-brane if they are
obtained by descending from X($=$A,B)-type for $SL(2,\br)$ part and
Y($=$A,B)-type for $SU(2)$ part.

First, let us consider the case of ${\rm Tr}\, g=2\kappa$ and
${\rm Tr}\, g'=2\xi$, namely the AA-brane. The conjugacy
classes are shifted by the gauge transformation \eqref{gauge} as
\begin{align}
  \cosh (\alpha + \beta + \tau + \rho) \cos \theta &= \kappa ,
 &\cos (\alpha ' + \beta ' + \tau + \rho) \cos \phi &= \xi .
\end{align}
Summing up the orbit we obtain
\begin{align}
 \alpha + \beta - \alpha ' - \beta '
 = \arccosh \left( \frac{\kappa}{\cos \theta} \right)
  -  \arccos \left( \frac{\xi}{\cos \phi} \right).
\end{align}
Therefore the geometry of the D-brane is given by the hypersurface
\begin{align}
 \lambda_+ - \lambda_+^0
 = \arccosh \left( \frac{\kappa}{\cos \theta} \right)
  -  \arccos \left( \frac{\xi}{\cos \phi} \right) .
  \label{D2-brane}
\end{align}
Because of the symmetry along $\lambda_+$, there is a freedom to
shift the total position by $\lambda_+^0$. Note that we have to sum up
$\lambda_+^0 = {\lambda_+^0}' + 2 \pi n$ with $0\leq
{\lambda_+^0}' < 2\pi , n \in {\mathbb Z}$ because of the residual
gauge transformation \cite{NW2}, which makes the D-branes travelling
in spiral orbits like D0-branes in Misner space (see appendix \ref{Misner}).
We should notice that $\cos \theta \leq \kappa$ and $\cos
\phi \geq |\xi|$. Therefore, there is no brane for $\kappa < 0$,
instantons at $\theta = \pi/2$ for $\kappa =0$, branes in the time period
$\arccos \kappa \leq \theta \leq \pi/2$ for $0 < \kappa < 1$, and
in all the time period $0 \leq \theta \leq \pi/2$ for $\kappa \geq 1$.
Similarly, there is no brane for $|\xi| > 1$, branes at
$0 \leq \phi \leq \arccos  |\xi|$ for $|\xi| < 1$ and lower dimensional
branes at $\phi = 0$ for $\xi = 1$.
 For $ \kappa > 0 $ and $|\xi| < 1$, the hypersurface  \eqref{D2-brane}
describes a D2-brane parametrized by $\theta,\phi$ and $\lambda_-$.
For $ \kappa > 0 $ and $|\xi| = 1$, the corresponding brane is D0-brane
at
\begin{align}
 \lambda_+ - \lambda_+^0
 = \arccosh \left( \frac{\kappa}{\cos \theta} \right)
  \label{D0-brane}
\end{align}
and $\phi = 0$. Notice that one more dimension is reduced because
a component of metric $g_{\lambda_- \lambda_-}$ is degenerated.

Next, let us move to the case of ${\rm Tr}\, g = 2 \kappa$ and
${\rm Tr}\, \sigma_1 g' = - 2 i \xi$, namely the AB-brane.
In this case, we sum up the gauge transformation \eqref{gauge} of
(twisted) conjugacy classes
\begin{align}
  \cosh (\alpha + \beta + \tau + \rho) \cos \theta &= \kappa ,
 &\sin (\alpha ' - \beta ' + \tau - \rho) \sin \phi &= \xi ,
\end{align}
which leads to the restriction of the worldvolume
\begin{align}
\cos \theta &\leq \kappa ,
&\sin \phi &\geq |\xi|.
\label{D3-brane}
\end{align}
As before, there is no brane for $\kappa <0$, instantons at
$\theta = \pi/2$ for $\kappa =0$, branes in $\theta \geq \arccos
\kappa$ for $0<\kappa<1$, and in all the time for $\kappa \geq 1$.
In the case of $\kappa > 0$, we have D3-branes for $|\xi| < 1$ and
D1-branes at $\phi = \pi/2$ for $|\xi| = 1$. This type of branes
resemble to D1-branes in Misner space explained in appendix
\ref{Misner}, though D1-branes in Misner space do not have the
ends.

Third, we consider the case of ${\rm Tr}\, \sigma _1 g = 2 \kappa$
and ${\rm Tr}\, g' = 2 \xi$, namely the BA-brane. In this case, we sum
up the gauge transformation \eqref{gauge} of (twisted) conjugacy classes
\begin{align}
  \sinh (\alpha - \beta + \tau - \rho) \sin \theta &= \kappa ,
 &\cos (\alpha ' + \beta ' + \tau + \rho) \cos \phi &= \xi ,
\end{align}
which leads to $\cos \phi \geq \xi$.
Notice that there is no restriction on the time $\theta$
contrary to the AB-brane case.
We have D3-brane for $|\xi| < 1$ and
D1-brane at $\phi = 0$ for $|\xi| = 1$.

Final case is ${\rm Tr}\, \sigma _1 g = 2 \kappa$
and ${\rm Tr}\, \sigma_1 g' = - 2 i \xi$, namely the BB-brane.
The hypersurface derived from the twisted conjugacy classes is
\begin{align}
 \lambda_- - \lambda_-^0
 = \arcsinh \left( \frac{\kappa}{\sin \theta} \right)
  -  \arcsin \left( \frac{\xi}{\sin \phi} \right) .
  \label{D2'-brane}
\end{align}
Again there is no restriction on the time $\theta$.
We have D2-brane satisfying \eqref{D2'-brane}
for $|\xi| <1$ and D0-brane at
\begin{align}
 \lambda_- - \lambda_-^0 + \frac{\pi}{2}
 = \arcsinh \left( \frac{\kappa}{\sin \theta} \right)
   \label{D0'-brane}
\end{align}
and $\phi = \pi /2$ for $|\xi| = 1$.

\subsection{Effective theories on D-branes}
\label{DBIregionII}

The purpose of this paper is to investigate the non-trivial
geometry of the NW model using the D-brane probes.
In general, D-branes and strings feel the background geometry
in different ways, which can be seen from their worldvolume actions.
The low energy actions for the worldvolume theories on D-branes are
of DBI-type, and D-brane metrics can be read from the
effective actions. In this subsection, we construct DBI actions
for the D-branes classified in the previous subsection, and see
how D-branes feel the background, in particular, near the singularities.
We also show that the classical solutions to the equations of motion
are consistent with the geometry of D-branes obtained above.

It is also important to examine how the non-trivial background affects
open string spectra on the D-branes.
A simple way to obtain the low energy spectra for open strings is
to utilize open string metrics \cite{SW}.
Because of non-trivial gauge flux on D-branes, open strings
feel the background metric in a modified way.
We denote induced closed string metric on D-brane as $g_{ab}$
and closed string coupling as $g_s$.
Then, open string metric $G_{ab}$
and open string coupling $G_s$ in a configuration with
${\cal F} = B + F$\footnote{We represent $B$ as induced $B$-field and
$F$ as gauge flux on D-brane.
We set for our convenience $\alpha' = \frac{1}{2\pi}$.}
are given by \cite{SW}
\begin{align}
 G_{ab} &= g_{ab} - {\cal F}_{ac} g^{cd} {\cal F}_{db} ,
 \label{ometric}\\
 G_s \equiv e^{\Phi_o}
 &= g_s \sqrt{\frac{- \det G}{ - \det (g + {\cal F})}} .
 \label{ocoupling}
\end{align}
The low energy spectra for open strings can be read from the
eigenfunctions of Laplacians in terms of open string quantities:
\begin{align}
 \Delta = \frac{1}{e^{- \Phi _o} \sqrt{-G}}
   \partial_a e^{- \Phi _o} \sqrt{-G} G^{ab} \partial_b .
   \label{olaplacian}
\end{align}
In this subsection, we only construct the Laplacians and see
their properties. More detailed analysis on wave functions is
given in section \ref{WF}.

\subsubsection{AA-brane}

Let us first examine the AA-type D0-brane with
$\kappa > 0$ and $|\xi| = 1$ ($\sin \phi=0$).
The DBI action in the static gauge is given by
\begin{align}
S & = -\tau _0 \int d \theta e^{-\Phi} \sqrt{-\det g} \nonumber \\
  &= - \tau_0 \int d \theta \sqrt{\sin ^2 \theta}
        \sqrt{1 - \cot ^2 \theta (\dot \lambda_+)^2}
   = - \tau_0 \int d \theta
        \sqrt{\sin ^2 \theta - \cos ^2 \theta (\dot \lambda_+)^2} .
\end{align}
In the second expression, the induced metric diverges due to
the singularity at $\theta = \phi = 0$. Fortunately,
the inverse of the string coupling vanishes at the point and
cancels the singularity as in the last expression.
Therefore, one can say that the D0-brane metric does not include
singularity at $\theta = \phi = 0$.\footnote{%
The similar phenomena is observed in the study of
D-branes in two dimensional Lorentzian black hole \cite{2DBHDbrane}.}

This action has translational symmetry along $\lambda_+$
direction, so the momentum conjugate to $\lambda_+$ is a constant
\begin{align}
 P_{\lambda_+} = \frac{\delta {\cal L}}{\delta (\dot \lambda_+)}
  = \frac{\tau_0 \cos ^2 \theta (\dot \lambda_+)}
         {\sqrt{\sin ^2 \theta - \cos ^2 \theta (\dot \lambda_+)^2}} ,
\end{align}
which can  be rewritten as
\begin{align}
 (\dot \lambda_+)^2 =
  \frac{\sin ^2 \theta}
  {\frac{\tau_0^2}{P^2_{\lambda_+}} \cos ^4\theta + \cos ^2\theta}.
\end{align}
On the other hand, from the equation \eqref{D0-brane}, one can
deduce
\begin{align}
  (\dot \lambda_+)^2 =
  \frac{\sin ^2 \theta}
  { - \frac{1}{\kappa^2} \cos ^4\theta + \cos ^2\theta} .
\end{align}
Thus the AA-type D0-brane traveling at
\eqref{D0-brane} has an imaginary momentum
$P^2_{\lambda_+} = - \kappa^2 \tau_0^2$, which means that
the D-brane travels faster than the speed of light.
In fact, the DBI action with the imaginary momentum is
\begin{align}
 S = - \tau_0 \int d \theta \sqrt{\frac{\sin ^2 \theta \cos ^2 \theta}
       {\cos ^2 \theta - \kappa ^2}} ,
\end{align}
which is also imaginary for $\cos \theta < \kappa$.
Therefore, we conclude that the D0-brane is tachyonic and is unphysical.%
\footnote{See \cite{BP} for discussions on the physicalness of D-branes in
$AdS_3$.} Here, we do not claim that there is no solution for
a D-brane with real momentum, but we just say that the
solution does not correspond to one of the D-branes classified
in the previous analysis.

Let us now move to AA-type D2-brane with the trajectory
\eqref{D2-brane}. In the static gauge, the DBI action is given by
\begin{align}
S \equiv \int d \theta d \phi d \lambda_- {\cal L} (\lambda_+,\dot A, A ')=
  - \tau_2 \int d \theta d \phi d \lambda _-  e^{-\Phi}
\sqrt{-\det(g + B+F)} . \label{DBI}
\end{align}
We only excite
$F_{\theta \lambda_-} = \partial_{\theta} A \equiv \dot A
(\theta,\phi)$ and $F_{\phi \lambda_-} = \partial_{\phi} A
\equiv A ' (\theta,\phi)$ in the gauge field strength,
then we have
\begin{align}
g + B + F =
  \begin{pmatrix}
   -1 + D \cot ^2 \phi (\dot \lambda _+)^2 &
   D \cot ^2 \phi \dot \lambda _+ \lambda ' _+ &
   - D \dot \lambda _+ + \dot A \\
   D \cot ^2 \phi \dot \lambda _+ \lambda ' _+ &
   1 + D \cot ^2 \phi (\lambda ' _+)^2 &
   - D \lambda '_+ + A ' \\
   D\dot \lambda _+ - \dot A &
   D \lambda '_+ - A ' &
   D \tan ^2 \theta
  \end{pmatrix} ,
\end{align}
where $D^{-1}=1 + \tan ^2 \theta \cot ^2 \phi$. Therefore, we find
\begin{align}
 &- \det (g+B+F) = D \tan ^2 \theta
 (1-D \cot ^2 \phi (\dot \lambda _+)^2)
 (1+D \cot ^2 \phi (\lambda ' _+)^2) \nonumber \\& \qquad
  + 2 D \cot ^2 \phi \dot \lambda _+ \lambda ' _+
    (- D \lambda '_+ + A ') (- D \dot \lambda _+ + \dot A)
  -  (1+D \cot ^2 \phi(\lambda ' _+)^2)
     (- D \dot \lambda _+ + \dot A)^2
   \nonumber \\& \qquad
  -  (- 1+D \cot ^2 \phi (\dot \lambda _+)^2)
     (- D \lambda ' _+ + A ')^2
  + D \tan ^2 \theta (D \cot ^2 \phi \dot \lambda _+ \lambda '_+)^2 .
\end{align}
The components of the field strength should satisfy the Gauss
constraints and $\lambda_+(\theta,\phi)$ should satisfy the
equation of motion;
\begin{align}
 \partial_{\theta}
 \left( \frac{\delta {\cal L}}{\delta \dot A} \right) &= 0 ,
 &\partial_{\phi}
 \left( \frac{\delta {\cal L}}{\delta A '} \right) &= 0 ,
 &\partial_{\theta}
 \left( \frac{\delta {\cal L}}{\delta \dot \lambda_+} \right)
 + \partial_{\phi}
 \left( \frac{\delta {\cal L}}{\delta \lambda '_+} \right) &= 0 .
 \label{gauss1}
\end{align}
Here we assume that
\begin{align}
 \dot A &= 0 , &A ' &= \lambda ' _+
 \label{ansatz1}
\end{align}
satisfy the above constraints \eqref{gauss1}, then the DBI action
\eqref{DBI} is written in a very simple form
\begin{align}
 S &= \int d \theta d \phi d \lambda _- {\cal L}' (\lambda_+)
  \nonumber \\
   &= - \tau_2 \int d \theta d \phi d \lambda _-
  \sqrt{(\sin ^2 \theta - \cos ^2 \theta (\dot \lambda _+)^2)
         (\sin ^2 \phi + \cos ^2 \phi (\lambda ' _+)^2)} .
\end{align}
Notice that the original metric has singularities at $\theta = \phi = 0$ and
$\theta = \phi = \pi /2 $, but these singularities
are cancelled by the contribution from the dilaton.
Using the ansatz \eqref{ansatz1}, the equations
\eqref{gauss1} can be rearranged as
\begin{align}
 \partial_{\theta}
 \left( \frac{\delta {\cal L}'}{\delta \dot \lambda_+} \right)
 + \partial_{\phi}
 \left( \frac{\delta {\cal L}'}{\delta \lambda '_+} \right) =
 \partial_{\theta}
 \left( \frac{\delta {\cal L}}{\delta \dot \lambda_+} \right)
 + \partial_{\phi}
 \left( \frac{\delta {\cal L}}{\delta \lambda '_+}
 +  \frac{\delta {\cal L}}{\delta A '} \frac{\delta A'}
       {\delta \lambda' _+} \right) = 0 .
       \label{eom1}
\end{align}
In other words, the problem is replaced by solving the equation of motion
with respect to the action ${\cal L}'$.
The equation of motion may reduce to
\begin{align}
 C_1 &= \frac{\cos ^2 \theta \dot \lambda_+}
   {\sqrt{\sin ^2 \theta - \cos ^2 \theta (\dot \lambda _+)^2 }} ,
&C_2 &= \frac{\cos ^2 \phi \lambda'_-}
   {\sqrt{\sin ^2 \phi + \cos ^2 \phi (\lambda ' _+)^2 }} ,
   \label{solution1}
\end{align}
or equivalently
\begin{align}
(\dot \lambda _+)^2 &= \frac{\sin ^2 \theta}
            {\frac{1}{C_1^2} \cos^4 \theta + \cos^2 \theta} ,
&(\lambda ' _+)^2 &= \frac{\sin ^2 \phi}
            {\frac{1}{C_2^2} \cos^4 \phi- \cos^2 \phi} .
\end{align}
Since we can show that
\begin{align}
 \frac{\delta {\cal L}}{\delta \dot A} &=  -
 \frac{\delta {\cal L}'}{\delta \dot \lambda_-},
 &\frac{\delta {\cal L}}{\delta A '} &=
 \frac{\delta {\cal L}'}{\delta \lambda '_-}
\end{align}
with the ansatz \eqref{ansatz1}, our choice of gauge fields satisfy
the Gauss constraints \eqref{gauss1} if we use the solution to
the equations \eqref{solution1}.
The solution to \eqref{solution1} is given by \eqref{D2-brane}
as expected if we assume $C_1^2 = - \kappa ^2$, $C_2^2 = \xi ^2$.%
\footnote{We suppose that the relative signs between $C_{1,2}$ and
$\kappa ,\xi $ are chosen appropriately.} However, with this
choice, the DBI action becomes
\begin{align}
 S = - \tau_2 \int d \theta d \phi d \lambda _-
  \sqrt{\frac{\sin ^2 \theta \cos ^2 \theta
              \sin ^2 \phi \cos ^2 \phi}
             {(\cos ^2 \theta - \kappa ^2)(\cos ^2 \phi - \xi ^2)}} .
\end{align}
The above action is purely imaginary because of the ranges of
$\theta,\phi$
($\cos \theta \leq \kappa,\cos \phi \geq |\xi|$).
This is again due to the tachyonic behavior of the D2-brane as in
the D0-brane case, and
we conclude that the AA-type D2-brane is also unphysical.

\subsubsection{AB-brane}

Again we start with a simpler case with $|\xi| = 1$ ($\cos \phi =
0$). The DBI action of the D-string wrapping along the $(\theta,
\lambda_-)$ plane is given by
\begin{align}
 S=-\tau_1 \int d \theta d \lambda_- e^{-\Phi} \sqrt{-\det (g+F)}
  =-\tau_1 \int d \theta d \lambda_-
   \sqrt{\sin ^2 \theta - \cos ^2 \theta f^2}
\end{align}
with $F_{\theta \lambda_-}=f$, where the singularity at
$\theta=\phi=\pi/2$ is cancelled as before.
The Gauss constraint leads to
\begin{align}
 \Pi = \frac{\delta {\cal L}}{\delta f}
     = \frac{\tau_1 \cos ^2 \theta f}
       {\sqrt{\sin ^2 \theta - \cos ^2 \theta f^2}}
\end{align}
with a constant $\Pi$,
which can also be written as
\begin{align}
 f^2 = \frac{\sin ^2 \theta}
  {\frac{\tau_1^2}{\Pi^2} \cos ^4\theta + \cos ^2\theta}.
\end{align}
If $\Pi ^2 = - \kappa ^2 \tau_1^2 $, then the condition $f^2 \geq
0$ is equivalent to $\cos \theta \leq \kappa$, which is consistent
with the previous results \eqref{D3-brane}.
However, the DBI action
\begin{align}
 S=-\tau_1 \int d \theta d \lambda_-
    \sqrt{\frac{\sin ^2 \theta \cos^2 \theta}{\cos^2 \theta - \kappa ^2}}
\end{align}
becomes imaginary also in this case.
This implies that the gauge field strength on the brane is
supercritical and the open strings on the brane are unstable
\cite{optical}.

We can analyze AB-type D3-brane with
$\kappa > 0, |\xi| < 1$ in a similar way.
We can show that the field strength
\begin{align}
 F_{\theta \lambda_-} &=0,
&F_{\phi \lambda_+} &=0,
&F_{\lambda_- \lambda_+} & = -1
\end{align}
satisfies the equations of motion.
Denoting
\begin{align}
 F_{\theta \phi} &= F (\theta, \phi),
&F_{\theta \lambda_+} &= \dot A_+ (\theta, \phi),
&F_{\phi \lambda_-} &= A'_- (\theta, \phi),
\end{align}
 we find
\begin{align}
 g+B+F =
  \begin{pmatrix}
   -1 & F & \dot A_- & 0 \\
   -F & 1 & 0 & A'_+  \\
    - \dot A_- & 0 & \bar D \tan ^2 \phi & - \bar D \\
   0 & - A'_+ & \bar D & \bar D \cot ^2 \theta
   \end{pmatrix},
\end{align}
where $\bar D^{-1} = 1+ \tan^2 \phi \cot ^2 \theta$. Thus the DBI
action is given by
\begin{align}
 S = - \tau_3 \int d \theta d \phi d \lambda_- d \lambda_+
       \sqrt{\sin ^2 \theta \cos ^2 \phi I }
\end{align}
with
\begin{align}
 I = 1 - (\dot A_-)^2 \cot ^2 \theta + (A'_+)^2 \tan ^2 \phi
           - F^2 - 2 \dot A_- A'_+ F - \bar D^{-1}(\dot A_-)^2(A'_+)^2 .
\end{align}
Since the solution to the equation $0=\frac{\delta {\cal L}}{\delta F}$
gives a consistent choice, we can set
\begin{align}
F = -  \dot A_- A'_+ .
\end{align}
Using this field strength, the DBI action is written as
\begin{align}
 S = - \tau_3 \int d \theta d \phi d \lambda_- d \lambda_+
       \sqrt{(\sin ^2 \theta - (\dot A_-)^2 \cos ^2 \theta)
             (\cos ^2 \phi + (A'_+)^2 \sin ^2 \phi)}.
\end{align}
At this stage, the singularities at $\theta=\phi=0,\pi/2$ are also
cancelled. The equations of motion for $A_-$ and $ A_+$
\begin{align}
 \partial_{\theta}
  \left(\frac{\delta {\cal L}}{\delta \dot A_-}\right) &= 0,
&\partial_{\phi}
  \left(\frac{\delta {\cal L}}{\delta \dot A_+}\right) &= 0
\end{align}
lead to the reduced equations
\begin{align}
C_1 &= \frac{\cos ^2 \theta\dot A_-}
     {\sqrt{\sin ^2 \theta  - (\dot A_-)^2 \cos ^2 \theta}} ,
&C_2 &= \frac{\sin ^2 \phi A'_+}
     {\sqrt{\cos ^2 \phi + (A'_+)^2 \sin ^2 \phi}}
\end{align}
with $C_{1}$ and $C_{2}$ being constants. These equations can be
rewritten as
\begin{align}
 (\dot A_-)^2 &= \frac{\sin ^2 \theta}
  {\frac{1}{C_1^2} \cos ^4 \theta + \cos ^2 \theta} ,
 &(A'_+)^2 &= \frac{\cos ^2 \phi}
  {\frac{1}{C_2^2} \sin ^4 \phi - \sin ^2 \phi} .
\end{align}
If we set $C_1^2 = - \kappa ^2,C_2^2 = \xi ^2 $, then $ (\dot
A_-)^2\geq 0, (A'_+)^2 \geq 0$ means $\cos ^2 \theta \leq
\kappa^2, \sin \phi > |\xi|$, which is the same condition as
\eqref{D3-brane}. However, with the above choice of field
strength, the DBI action reduces to
\begin{align}
 S = - \tau_3 \int d \theta d\phi d \lambda_- d \lambda_+
       \sqrt{\frac{\sin^2 \theta \cos ^2 \theta
                           \sin^2 \phi\cos ^2 \phi}
            {(\cos ^2 \theta - \kappa^2)(\sin ^2 \phi - \xi^2)}} ,
\end{align}
which is purely imaginary in the region \eqref{D3-brane}. Therefore,
we conclude that the AB-type D3-brane is unphysical because of the
presence of supercritical gauge flux.

\subsubsection{BA-brane}

In the case of BA-type D1-brane ($|\xi| = 1$), the
DBI action is given by
\begin{align}
 S = - \tau_1 \int d \theta d \lambda_+ e^{-\Phi} \sqrt{- \det (g + F)}
   = - \tau_1 \int d \theta d \lambda_+
    \sqrt{\cos ^2 \theta - \sin ^2 \theta f^2 },
\end{align}
where we set $F_{\theta \lambda_+} = f$.
The singularity at $\theta = 0$ is cancelled as before.
The Gauss constraint leads to
\begin{align}
 \frac{ \delta {\cal L}}{\delta f}
   = \frac{\tau_1 \sin ^2 \theta f}
     {\sqrt{\cos ^2 \theta - \sin ^2 \theta f^2 }} = \Pi
\end{align}
with a constant $\Pi$, and one can solve for $f$ as
\begin{align}
 f^2 = \frac{\cos^2 \theta}
 {\frac{\tau_1^2 }{\Pi^2}\sin^4 \theta + \sin^2 \theta} .
 \label{BAfluxII}
\end{align}
For all $\theta$, $f^2$ is non-negative, and the DBI
action on the D-string is given by
\begin{align}
 S = - \tau_1 \int d \theta d \lambda_+ \cos \theta
 \sqrt{\frac{\sin^2 \theta }{\sin^2 \theta + \frac{\Pi^2}{\tau_1^2}}} ,
\end{align}
which is real. Therefore, one expects that the resulting D-brane is physical one,
opposite to the previous cases.

Next let us consider BA-type D3-brane with general $|\xi| < 1$. We
excite the following components of gauge field strength
\begin{align}
 F_{\theta \phi} &= F ,
&F_{\theta \lambda_+} &= \dot A_+ ,
&F_{\phi \lambda_-} &= A'_- ,
\end{align}
then the DBI action  becomes
\begin{align}
 S = - \tau_3 \int d \theta d \phi d \lambda_- d \lambda_+
       \sqrt{\cos ^2 \theta \sin ^2 \phi I }
\end{align}
with
\begin{align}
 I = 1 - (\dot A_+)^2 \tan ^2 \theta + (A'_-)^2 \cot ^2 \phi
           - F^2 - 2 \dot A_+ A'_- F - D^{-1}(\dot A_+)^2(A'_-)^2 .
\end{align}
Inserting $F = -  \dot A_+ A'_-$, which satisfies
$\frac{\delta {\cal L}}{\delta F}=0$, into the DBI action, we obtain
\begin{align}
 S = - \tau_3 \int d \theta d \phi d \lambda_- d \lambda_+
       \sqrt{(\cos ^2 \theta - (\dot A_+)^2 \sin ^2 \theta)
             (\sin ^2 \phi + (A'_-)^2 \cos ^2 \phi)}.
\end{align}
As before, there is no singularity at $\theta=\phi=0,\pi/2$ in this action.
The equations of motion for $A_+, A_-$ are given by
\begin{align}
 \partial_{\theta}
  \left(\frac{\delta {\cal L}}{\delta \dot A_+}\right) &= 0,
&\partial_{\phi}
  \left(\frac{\delta {\cal L}}{\delta \dot A_-}\right) &= 0,
\end{align}
hence we may have
\begin{align}
C_1 &= \frac{\sin ^2 \theta\dot A_+}
     {\sqrt{\cos ^2 \theta  - (\dot A_+)^2 \sin ^2 \theta}} ,
&C_2 &= \frac{\cos ^2 \phi A'_-}
     {\sqrt{\sin ^2 \phi + (A'_-)^2 \cos ^2 \phi}}
\end{align}
with constants $C_{1}$ and $C_{2}$, or inversely
\begin{align}
 (\dot A_+)^2 &= \frac{\cos ^2 \theta}
  {\frac{1}{C_1^2} \sin ^4 \theta + \sin ^2 \theta} ,
 &(A'_-)^2 &= \frac{\sin ^2 \phi}
  {\frac{1}{C_2^2} \cos ^4 \phi - \cos ^2 \phi} .
\end{align}
If we set $C_1^2= \kappa^2, C_2^2 = \xi^2$,\footnote{Actually, the
relation between $C_1$ and $\kappa$ is not fixed uniquely in this way
of analysis. We might be able to determine it by gauging the case of
$SL(2,\br) \times SU(2)$ WZW model \cite{Sarkissian}.} then the conditions
$(\dot A_+)^2 \geq 0$ and $(A'_-)^2 \geq 0$ imply $\cos \phi \geq |\xi|$,
which is the same as the condition obtained before.
With the field strength, the DBI action becomes
\begin{align}
 S = - \tau_3 \int d \theta d\phi d \lambda_- d \lambda_+
       \sqrt{\frac{\sin^2 \theta \cos ^2 \theta
                           \sin^2 \phi\cos ^2 \phi}
            {(\sin ^2 \theta + C_1^2)(\cos ^2 \phi -C_2^2)}} .
\end{align}
It is real for $\cos \phi \geq |\xi|$, and hence the D3-brane in
this case is physical.
Since we now obtain the physical D-brane,
let us examine the spectrum for open strings on the D3-brane.

As we mentioned before, wave functions are given by the
eigenfunctions of Laplacian expressed in terms of open string quantities.
The open string metric and the open string coupling with
our configuration of gauge fields are computed by following
\eqref{ometric} and \eqref{ocoupling} as
\begin{align}
 ds^2_{open}& = - \alpha {d\theta}^2 + \beta {d\phi} ^2
  + \alpha \cot ^2 \theta  (d \lambda_+ - A '_- d {\phi})^2
  + \beta \tan ^2 {\phi}(d \lambda_- + \dot A _+ d \theta )^2 ,\\
  G_s &= \sqrt{\frac{\alpha \beta}{\sin ^2 \theta \cos ^2 {\phi}}},
\end{align}
with
\begin{align}
 \alpha &= 1 - \tan ^2 \theta (\dot A_+)^2
         = \frac{\sin ^2 \theta}{\sin ^2 \theta + C_1^2},
&\beta & = 1 + \cot ^2 {\phi}(A_- ')^2
         = \frac{\cos ^2 {\phi}}{\cos ^2 {\phi} - C_2^2} .
\end{align}
Let us make the transformations
\begin{align}
 \lambda_+ - \int^{{\phi}} A '_- (x) dx &\to \lambda_+,
 &\lambda_- + \int^{\theta } \dot A_+ (x) dx &\to \lambda_- ,
\end{align}
and
\begin{align}
 \cos \theta &\to \sqrt{1 + C_1^2} \cos \theta ,
 &\sin {\phi} &\to \sqrt{1-C_2^2} \sin {\phi} ,
 \label{BAIIrescale}
\end{align}
then the open string metric and the open string coupling
become
\begin{align}
 ds_{open}^2 &= - {d\theta}^2 + {d\phi} ^2
 + \cot ^2 \theta d  \lambda_+ ^2 + \tan ^2 {\phi} d  \lambda_- ^2 ,\\
 G_s &=\frac{1}{ \sqrt{(1+C_1^2)(1-C_2^2) \sin^2 \theta \cos^2 {\phi}}}.
\end{align}
We should also note that the region of $\theta$ is restricted as
$\cos ^2 \theta \leq 1/(1+C_1^2)$. The  Laplacian in terms of open
string metric can be written as
\begin{align}
 \Delta = - \frac{1}{\cos \theta} \partial_{\theta}
                     \cos \theta \partial_{\theta}
          + \frac{1}{\sin {\phi}} \partial_{{\phi}}
                     \sin {\phi} \partial_{{\phi}}
          + \tan ^2 \theta \partial_{\lambda_+}^2
          + \cot ^2 {\phi} \partial_{\lambda_-}^2 .
          \label{laplacianBAII}
\end{align}
In this expression, we can observe that not only the DBI action
but also wave functions on the D-brane do not behave in a
peculiar way at the singularities $\theta = \phi = 0,\pi/2$.
We continue the analysis on the wave function in section \ref{WF}.

We can also see that the expression of Laplacian is consistent
with the boundary conditions of currents.
If we denote $J^A$ and $K^A$ ($\bar J^A$ and
$\bar K^A$) $( A = +, -, 3)$
as the holomorphic (anti-holomorphic) part of currents
in $SL(2,\mathbb{R})$ and $SU(2)$ WZW models, then the boundary
conditions in this case are given by
\begin{align}
 J^3 &= - \bar J^3 , &J^{\pm} &= \bar J^{\mp} ,
&K^3 &= \bar K^3 , &K^{\pm} &= \bar K^{\pm}
\end{align}
at the boundary of worldsheet. The remaining symmetries generated
by $J^3 - \bar J^3$ and $K^3 + \bar K^3$  correspond to the
translations of $\lambda_+$ and $\lambda_-$, respectively, which is
consistent with the form of the Laplacian \eqref{laplacianBAII}.
We should notice that the Laplacian (or its
eigenfunction) is quite simple even though the background metric is
rather complicated. This is due to the coset construction.

\subsubsection{BB-brane}

For BB-type D0-brane, DBI action becomes
\begin{align}
 S = - \tau_0 \int d \theta e^{-\Phi} \sqrt{- \det g}
   = - \tau_0 \int d \theta
    \sqrt{\cos ^2 \theta - \sin ^2 \theta \dot \lambda_-^2 } ,
\end{align}
because of the condition $\cos \phi = 0$. The momentum conjugate
to $\lambda_-$ is a constant of motion
\begin{align}
 P_{\lambda_-} = \frac{\delta {\cal L}}{\delta \dot \lambda_-}
  = \frac{\tau_0 \sin ^2 \theta \dot \lambda_- }
         {\sqrt{\cos^2 \theta - \sin ^2 \theta \dot \lambda _-^2}} ,
\end{align}
therefore
\begin{align}
 \dot \lambda _-^2 = \frac{\cos ^2 \theta}
 {\frac{\tau^2 _0}{{P_{\lambda _-}^2}}
  \sin ^4 \theta + \sin ^2\theta },
\end{align}
which is consistent with the motion of D0-brane \eqref{D0'-brane}
with $P^2_{\lambda _-} / \tau^2 _0 = \kappa ^2$. We can show that the
DBI action
\begin{align}
 S = - \tau_0 \int d \theta
    \sqrt{\frac{\sin ^2 \theta \cos ^2 \theta}
         {\sin ^2 \theta + \frac{P^2_{\lambda_-}}{\tau_0}}}
\end{align}
is real and the corresponding D0-brane travels at a speed
less than that of light.
We should remark that even if the physical D0-brane is used
as a probe, the singularity at $\theta = \phi = \pi /2$
does not appear in the D-brane metric as before.
One might expect that the point particle feels the singularity
more severely than string, but actually what we have shown is that
the D0-brane feel the singularity in a milder way than string.

BB-type D2-brane can be analyzed in a way similar to
the AA-type D2-brane.
Here we only introduce non-trivial potential
$A_{\lambda_+} = A(\theta,\phi)$
and also assume that the gauge field
\begin{align}
 \dot A &= - \dot \lambda_- , &A ' &= 0
 \label{ansatzII}
\end{align}
satisfies the equations of motion.
Then the DBI action becomes very simple form as
\begin{align}
 S = - \tau_2 \int d \theta d \phi d \lambda _+
  \sqrt{(\cos ^2 \theta - \sin ^2 \theta (\dot \lambda _-)^2)
         (\cos ^2 \phi + \sin ^2 \phi (\lambda ' _-)^2)} .
\end{align}
As before, we can use this action to derive the equations of
motion for $\lambda_-$, which reduce to
\begin{align}
 C_1 &= \frac{\sin ^2 \theta \dot \lambda_-}
   {\sqrt{\cos ^2 \theta - \sin ^2 \theta (\dot \lambda _-)^2 }} ,
&C_2 &= \frac{\sin ^2 \phi \lambda'_-}
   {\sqrt{\cos ^2 \phi + \sin ^2 \phi (\lambda ' _-)^2 }} .
\end{align}
We can also show that the gauge fields \eqref{ansatzII}
satisfy the Gauss constraints from these equations.
Because the above equations can be rewritten as
\begin{align}
(\dot \lambda _-)^2 &= \frac{\cos ^2 \theta}
            {\frac{1}{C_1^2} \sin^4 \theta + \sin^2 \theta} ,
&(\lambda ' _-)^2 &= \frac{\cos ^2 \phi}
            {\frac{1}{C_2^2} \sin^4 \phi- \sin^2 \phi} ,
\end{align}
we can reproduce the geodesic of D2-brane \eqref{D2'-brane}
if we assume $C_1^2 = \kappa^2$, $C_2^2 = \xi^2$.
The DBI action
\begin{align}
 S = - \tau_2 \int d \theta d \phi d \lambda _+
  \sqrt{\frac{\sin ^2 \theta \cos ^2 \theta
              \sin ^2 \phi \cos ^2 \phi}
             {(\sin ^2 \theta + \kappa ^2 )(\sin ^2 \phi - \xi ^2)}}
\end{align}
is real for $\sin \phi \geq |\xi|$, thus we conclude that
the BB-type D2-brane is also physical.
Notice that the D-brane does not seem to feel the singularity at
$\theta = \phi = \pi/2$ even for the physical D-brane
with higher dimensionality.

The open string metric \eqref{ometric} and the open string coupling
\eqref{ocoupling} in this case are computed as
\begin{align}
 ds^2_{open} &= - \alpha {d\theta}^2 + \beta {d\phi} ^2
  + \frac{\beta \cot ^2 {\phi} }
  {1 + \alpha^{-1} \tan ^2 \theta \beta \cot^2 {\phi} }
   d\lambda^2_+,\\
 G_s &= \frac{1}{\sqrt{\cos ^2 \theta \sin ^2 {\phi}
    (\beta ^{-1} + \tan^2 \theta \cot^2 {\phi} \alpha ^{-1})}} ,
\end{align}
where
\begin{align}
 \alpha &= 1 - \tan ^2 \theta (\dot \lambda_-)^2
         = \frac{\sin ^2 \theta}{\sin ^2 \theta + C_1^2},
&\beta & = 1 + \tan ^2 {\phi}(\lambda_- ')^2
         = \frac{\sin ^2 {\phi}}{\sin ^2 {\phi} - C_2^2} .
\end{align}
Performing the transformations
\begin{align}
 \cos \theta &\to \sqrt{1 + C_1^2} \cos \theta ,
 &\cos {\phi} &\to \sqrt{1 - C_2^2} \cos {\phi} ,
\end{align}
the open string metric and the string coupling become
\begin{align}
 ds^2_{open} &= - {d\theta}^2 + {d\phi} ^2
  + \frac{\cot ^2 {\phi}}
  {1 + \tan ^2 \theta \cot^2 {\phi} } d\lambda^2_+ ,\\
 G_s &= \frac{1}{\sqrt{(1+C_1^2)(1-C_2^2)
    (\cos ^2 \theta \sin ^2 {\phi}
   + \sin ^2 \theta \cos ^2 {\phi})}} .
\end{align}
In terms of the redefined parameters, the Laplacian can be written
as
\begin{align}
 \Delta = - \frac{1}{\cos \theta} \partial_{\theta}
            \cos \theta \partial_{\theta}
          + \frac{1}{\cos {\phi}} \partial_{{\phi}}
            \cos {\phi}\partial_{{\phi}}
          + (\tan ^2 \theta + \tan ^2 {\phi})
            \partial_{\lambda_+}^2 .
            \label{laplacianBBII}
\end{align}
It might be interesting to notice that the Laplacians for BA-brane
and for BB-brane are quite similar even though
the open string metrics are rather different.
This is again due to the coset construction.
The boundary conditions for the currents in this case are
\begin{align}
 J^3 &= - \bar J^3 , &J^{\pm} &= \bar J^{\mp} ,
&K^3 &= - \bar K^3 , &K^{\pm} &= \bar K^{\mp} ,
\end{align}
and the remaining symmetries generated by $J^3 - \bar J^3$
and $K^3 - \bar K^3$ corresponds to the shift of
$\lambda_+$. This is again consistent with the form of the Laplacian
\eqref{laplacianBBII}.

%%%%%%%%%%%%%%%%%%%%%%%%%%%%%%%%%%%%%%%%%%%%%%%%%%%%%%%%%%%%%%%%%%%%%%%%%%%

\section{D-branes in the whisker regions}
\label{D-branew}

The whisker regions contain CTCs,
and they are usually regarded as sources of pathologies.
However, since the Nappi-Witten model is constructed as a gauged WZW model,
it seems natural to think that the model describes a consistent
background. A famous example with non-trivial CTC is G\"{o}del
universe \cite{Godel}, and recently there are several studies,
e.g., in \cite{GGHPR,BGHV,HT,DFS1,HR,BDPR,BHH,JR} regarding the
properties of CTCs in superstring backgrounds of  G\"{o}del
type. In particular, ref. \cite{DFS1} studied the regions including
CTCs by using a D-brane wrapping the CTC as a probe.
In this section, we investigate the whisker regions (region 1 and
region $2'$) by using D-branes as probes with a special
care on CTCs and related pathologies.
In particular, we examine wave functions on D-branes wrapping the
CTCs.

\subsection{Geometry of D-branes: A group theoretic view}
\label{conjugacyw}

The geometry of D-branes can be read off from the
conjugacy classes and twisted ones of $SL(2,\br)$ and $SU(2)$ as in region II.
In region 1, the $SL(2,\br)$ part is changed to
\begin{align}
 &{\rm A}:\quad
  {\rm Tr}\, g = 2 \cosh (\alpha + \beta) \cosh \theta = 2 \kappa ,\\
 &{\rm B}:\quad
 {\rm Tr}\, \sigma_1 g = 2 \cosh (\alpha - \beta) \sinh \theta = 2 \kappa ,
\end{align}
but the $SU(2)$ part is the same as
\eqref{SU2A} and \eqref{SU2B}.
The (twisted) conjugacy class of $SL(2,\br)$ in region $2'$ is
\begin{align}
 &{\rm A}:\quad
  {\rm Tr}\, g = 2 \sinh (\alpha + \beta) \sinh \theta = 2 \kappa ,\\
 &{\rm B}:\quad
 {\rm Tr}\, \sigma_1 g = 2 \sinh (\alpha - \beta) \cosh \theta = 2 \kappa ,
\end{align}
and that of $SU(2)$ is
\begin{align}
 &{\rm A}:\quad
 {\rm Tr}\, g' = - 2 \cos (\alpha ' + \beta ') \sin \phi = 2 \xi ,\\
 &{\rm B}:\quad
 {\rm Tr}\, \sigma_1 g'= - 2 i \sin (\alpha ' - \beta ') \cos \phi=
   - 2 i \xi ,
\end{align}
because of the shift in $\phi:~\phi \to \phi + \pi /2$.
As before, we obtain the geometry of D-branes in the coset theory
by summing over the gauge orbit of the above (twisted) conjugacy
classes, projecting into the gauge invariant space, and shifting by
$U(1) \times U(1)$.

Let us now consider the brane trajectory in region 1. Following
the analysis of region II, we can see that AA-branes are described
by the hypersurface
\begin{align}
 \lambda_+ - \lambda_+^0
 = \arccosh \left( \frac{\kappa}{\cosh \theta} \right)
  -  \arccos \left( \frac{\xi}{\cos \phi} \right) .
  \label{D2-brane1}
\end{align}
As we mentioned before, the time coordinate is either $\lambda_+$
or $\lambda_-$ in the whisker regions, so we use $\phi, \lambda_{\pm}$
as the coordinates of the worldvolume of the brane. Suppose $\kappa
> 1$, then we have D2-brane for $|\xi| < 1$ and D0-brane at $\phi=
0$ for $|\xi| = 1$. For $\kappa = 1$, we have lower dimensional
D-branes. If we regard $\lambda_+$ as the time coordinate, then
the D2-brane behaves in a way similar to D0-brane in a
whisker region of Misner space (see appendix \ref{Misner}). Now
that $\lambda_-$ does not depend on the other coordinates
in \eqref{D2-brane1}, the D2-brane wraps the CTC in the region
where $\lambda_-$ takes the role of time.

Similarly, we can see that AB-brane has the worldvolume bounded by
\begin{align}
 \cosh \theta &\leq \kappa ,
&\sin \phi &\geq |\xi| .
\label{D3-braneAB1}
\end{align}
Therefore we have D3-brane for $\kappa > 1$ and $|\xi| < 1$,
D2-instanton at $\theta = 0$ labeled by $\phi, \lambda_+$
for $\kappa = 1$ and $|\xi| < 1$,
D1-brane at $\phi = \pi/2$ labeled by $\theta , \lambda_-$
for $\kappa > 1$ and $|\xi| =  1$.
Since $\lambda_{\pm}$ are independent on the other coordinates
in \eqref{D3-braneAB1}, the worldvolume of the D3-brane
includes CTC everywhere in the region \eqref{D3-braneAB1}.

The worldvolume of BA-brane is shown to be
\begin{align}
\sinh &\theta \leq \kappa ,
&\cos &\phi \geq |\xi| .
\label{D3-braneBA1}
\end{align}
Therefore, we have D3-brane for $\kappa  > 0, |\xi| <1$,
D1-brane labeled by $\theta,\lambda_+$ at $\phi = 0$
for $\kappa > 0,|\xi| = 1$, and
D2-instanton labeled by $\theta, \lambda_+$ at $\theta = 0$
for $\kappa = 0, |\xi| <1 $.
The D3-brane also includes CTC on it everywhere
in the region \eqref{D3-braneBA1}.

For BB-brane, the geometry is obtained as
\begin{align}
 \lambda_- - \lambda_-^0
 = \arccosh \left( \frac{\kappa}{\sinh \theta} \right)
  -  \arcsin \left( \frac{\xi}{\sin \phi} \right) .
  \label{D2'-brane1}
\end{align}
Therefore, supposing $\kappa > 0$, we have D2-brane satisfying
\eqref{D2'-brane1} for $|\xi| < 1$ and
D0-brane labeled by $\lambda_-$ at $\phi = \pi /2$ for $|\xi| = 1$.
Just like the AA-type D2-brane, the BB-type D2-brane includes CTC on it
in the region where $\lambda_+$ is regarded as the time.

The region $2'$ can be analyzed in a similar
way. AA-brane is described by the hypersurface
\begin{align}
 \lambda_+ - \lambda_+^0 =
 \arcsinh \left(\frac{\kappa}{\sinh \theta}\right)
 - \arccos \left( \frac{-\xi}{\sin \phi} \right) ,
 \label{D2-brane2}
\end{align}
so we have D2-brane for $|\xi| < 1$ and D0-brane for $|\xi| = 1$.
For AB-brane, we have D3-brane in the region where
$\cos \phi \geq |\xi|$ for $|\xi| <1$ and D1-brane at $\phi = 0$
for $|\xi| = 1$. For BA-brane, we have D3-brane in the
region where $\sin \phi \leq - |\xi|$ for $|\xi| <1$ and
D1-brane at $\phi = -\pi/2$ for $|\xi| = 1$.  Similarly one can
show that for BB-brane there is D2-brane at the
hypersurface
\begin{align}
 \lambda_- - \lambda_-^0 =
 \arcsinh \left(\frac{\kappa}{\cosh \theta}\right)
 - \arcsin \left( \frac{-\xi}{\cos \phi} \right) ,
  \label{D2'-brane2}
\end{align}
for $|\xi| <1$ and D0-brane for $|\xi| = 1$. Note that for all
the branes $\theta$ runs up to infinity ($\theta = \infty$), where
the hypersurface may take the role of boundary of AdS space
\cite{NW2}. Other properties, such as the similarity with the
branes in Misner space and the inclusion of CTC, are very similar to
those of region 1.

\subsection{Effective actions on D-branes in region 1}
\label{DBIregion1}

As we have seen in the previous subsection, D-branes in region 1 extend
up to finite $\theta$, but D-branes in region 2 extend all the
way to the ``boundary'' at $\theta = \infty$. It is natural to
expect that these two types of D-branes are qualitatively
different.
In this section, we construct DBI actions for effective theories
on D-branes in region 1 and study the wave functions on them as in
the previous section.

\subsubsection{AA-brane}

Let us first examine the D0-brane with $\kappa > 1$ and $|\xi| =
1$, namely with $\sin \phi = 0$. In this case, the $g_{\lambda_+
\lambda_+}$ component of the metric is negative, so we set $\lambda_+$ to be the
time-coordinate of the worldvolume, as a gauge choice. Then the
DBI action of the D0-brane is given by
\begin{align}
S & = -\tau _0 \int d \lambda_+ e^{-\Phi} \sqrt{-\det g} \nonumber \\
  & = - \tau_0 \int d \lambda_+
        \sqrt{\cosh ^2 \theta - \sinh ^2 \theta (\dot \theta)^2} ,
\end{align}
where we denote $\dot{} = d/d \lambda_+$. The singularity of the
metric at $\theta = 0$ is once again cancelled with the inverse of the
string coupling. Since the metric does not depend on the time
$\lambda_+$, the energy is a constant
\begin{align}
 E = \frac{\delta {\cal L}}{\delta (\dot \theta)} \dot \theta - {\cal L}
  = \frac{\tau_0 \cosh ^2 \theta }
         {\sqrt{\cosh ^2 \theta - \sinh ^2 \theta (\dot \theta)^2}}.
\end{align}
This equation can be rewritten as
\begin{align}
 (\dot \theta)^2 =
  \frac{\cosh ^2 \theta - \frac{\tau_0^2}{E^2} \cosh ^4 \theta}
  {\sinh ^2 \theta} ,
\end{align}
which is consistent with the geometry \eqref{D2-brane1} if we set
$E^2/\tau_0^2 = \kappa^2$. The DBI action with the above choice is
given by
\begin{align}
 S = - \tau_0 \int d \lambda_+ \frac{\cosh ^2 \theta}{\kappa} .
\end{align}
Notice that the action is real, and hence the D-brane should be
physical.

Next, we analyze the D2-brane of this type. We set the worldvolume
coordinates as $(\lambda_+,\lambda_-,\phi)$ such that one of them
behaves like time-coordinate. The DBI action is now written as
\begin{align}
S \equiv \int d\lambda_+ d\lambda_- d\phi {\cal L}
  = -\tau_2 \int d\lambda_+ d\lambda_- d\phi e^{-\Phi}
\sqrt{-\det (g + B + F )}  .
\end{align}
Let us excite only $F_{\lambda_- \phi}=F$, then we obtain
\begin{align}
 g + B + F =
  \begin{pmatrix}
   D \cot ^2 \phi + ( \dot \theta )^2 &
   D &\dot \theta \theta ' \\
   - D & -D \tanh ^2 \theta & F \\
   \dot \theta \theta' & -F & 1 + (\theta ')^2
  \end{pmatrix} ,
\end{align}
where we denote $ {}' = d/d\phi$ and $D^{-1}= 1 - \tanh ^2 \theta
\cot ^2 \phi $.
Now we have
\begin{align}
-\det (g+B+F) = - D(1+(\theta ')^2) + D\tanh ^2 \theta (\dot \theta)^2
 - 2DF\dot \theta \theta ' - F^2 (D\cot ^2 \phi + (\dot\theta)^2).
\end{align}
In order to find classical configuration, we can utilize the Gauss
constraints and the conservation of the energy momentum tensor:
\begin{align}
 \partial_{\phi} \left(\frac{\delta \cal L}{\delta F} \right) &=0 ,
 &\partial_{\lambda_+} T^{\lambda_+}_{ ~ \lambda_+}
  + \partial_{\phi} T^{\phi}_{ ~ \lambda_+} &= 0 ,
\end{align}
where the energy momentum tensor is defined as
$(\mu,\nu,\rho=\lambda_{\pm},\phi)$
\begin{align}
 T^{\mu}_{ ~ \nu} =
 \frac{\delta {\cal L}}{\delta \partial_{\mu} \theta} \partial_{\nu} \theta
 +  \frac{\delta {\cal L}}{\delta F_{\mu \rho}} F_{\rho \nu}
 - \delta^{\mu}_{ ~ \nu} {\cal L}.
\end{align}

Once we assume that $F=- \theta '/\dot \theta$, the DBI action
takes a simpler form
\begin{align}
 S = - \tau_2 \int d \lambda_+ d \lambda _- d \phi
  \sqrt{- {\rm sgn} D (\cosh ^2 \theta - \sinh ^2 \theta
    (\dot \theta )^2)
         \left(\sin ^2 \phi + \cos ^2 \phi
   \frac{{(\theta ')}^2}{(\dot \theta)^2}\right)},
\end{align}
where ${\rm sgn} D = 1$ for $D < 0$ and ${\rm sgn} D = -1$ for $D
> 0$. With the above choice of $F$ $(F=- \theta '/\dot \theta)$, the Gauss constraint leads to
\begin{align}
 \frac{\frac{\theta '}{\dot \theta} \cos ^2 \phi}
      {\sqrt{\sin ^2 \phi + \cos ^2 \phi
   \frac{{(\theta ')}^2}{(\dot \theta)^2}}} = C ,
\end{align}
which can be rewritten as
\begin{align}
 \frac{{(\theta ')}^2}{(\dot \theta)^2}
  = \frac{\sin ^2 \phi}
    {\frac{1}{C^2} \cos ^4 \phi - \cos ^2 \phi }
    \label{geometryAA11}
\end{align}
with a constant $C$. Therefore, $F=-\theta ' /\dot \theta$ satisfies the
Gauss constraint if $\theta$ satisfies \eqref{geometryAA11}.
The energy momentum tensor with $F=- \theta '/\dot \theta$ is given by
\begin{align}
 T^{\lambda_+}_{~ \lambda_+} &
 = \tau_2 \sqrt{\sin ^2 \phi + \cos ^2 \phi
   \frac{{(\theta ')}^2}{(\dot \theta)^2}}
   \frac{\cosh ^2 \theta}{\sqrt{ - {\rm sgn} D (\cosh ^2 \theta
          - \sinh ^2 \theta (\dot \theta)^2 )}},
 &T^{\phi}_{~ \lambda_+} &= 0 ,
\end{align}
thus the configuration with
\begin{align}
 E = \frac{\cosh ^2 \theta}{\sqrt{ - {\rm sgn} D (\cosh ^2 \theta
          - \sinh ^2 \theta (\dot \theta)^2 ) }},
\end{align}
or
\begin{align}
 (\dot \theta)^2 =
 \frac{\cosh ^2 \theta + \frac{{\rm sgn} D}{E^2} \cosh ^4 \theta}
      {\sinh ^2 \theta}
      \label{geometryAA12}
\end{align}
satisfies the conservation of the energy momentum tensor.

Note that if we assign $C^2 = \xi^2$ and $E^2 = - {\rm sgn} D
\kappa ^2$, then the solution to \eqref{geometryAA11} and
\eqref{geometryAA12} is indeed \eqref{D2-brane1}, which was
obtained by utilizing the conjugacy classes. The DBI action with
these fields becomes
\begin{align}
 S = - \tau_2 \int d \lambda_+ d \lambda_- d \phi
  \sqrt{\frac{\cosh ^4 \theta
              \sin ^2 \phi \cos ^2 \phi}
             { - {\rm sgn} D \kappa ^2 (\cos ^2 \phi - \xi ^2 ) }} ,
\end{align}
which is real for $D < 0$ and imaginary for $D > 0$. Therefore,
the condition for the D-brane to be physical is changed across the
line $D=0$. As mentioned before, the time and space coordinates
$\lambda_{+}$ and $\lambda_-$ are exchanged across the line $D=0$
\eqref{singular-surface}, so the change of physicalness
is natural from the bulk viewpoint.
However, in the DBI action, the singularity from the
metric is canceled with the divergence of the string coupling, and
only the sign is left.
Thus, those who live on the brane might feel strangely the existence of
the domain wall at $D=0$.
For the physical region $D < 0$, $\lambda_+$ serves as a
time-coordinate, and hence CTC is not included on the brane.
As we will see below, this happens not for all the branes.

Let us move to the equations for wave functions on the D2-brane,
where only the physical region $D < 0$ is considered.
The open string metric has been computed as
\begin{align}
 ds_{open}^2 &= \beta {d\phi} ^2 -
 \alpha (\dot \theta d \lambda_+ + \theta ' d {\phi})^2
  + \frac{\alpha \beta \tanh ^2 \theta}
  {\beta + \alpha \tanh ^2 \theta \cot^2 {\phi} }
   d\lambda^2_- ,
\end{align}
with
\begin{align}
 \alpha &= - 1 + \coth ^2 \theta \frac{1}{(\dot \theta)^2}
         = \frac{\cosh ^2 \theta}{E^2 - \cosh ^2 \theta},
&\beta & = 1 + \cot ^2 {\phi}\frac{{(\theta ')}^2}{(\dot
\theta)^2}
         = \frac{\cos ^2 {\phi}}{\cos ^2 {\phi} - C^2} .
\end{align}
Changing the coordinate system from $(\lambda_+, \lambda_-,
{\phi})$ into $(\lambda_-, \theta, {\phi})$, the metric is
rewritten as
\begin{align}
 ds_{open}^2 &= - \alpha {d\theta}^2 + \beta {d\phi} ^2
  + \frac{\alpha \beta \tanh ^2 \theta}
  {\beta + \alpha \tanh ^2 \theta \cot^2 {\phi} }
   d\lambda^2_- .
\end{align}
It is amusing to note that the time-coordinate in the open string
metric is now $\theta$, which was a space-like coordinate in the
closed string metric. A person living on the brane seems to feel
that time $\theta$ is running from $0$ to $\arccosh E$. Taking
care of the Jacobian due to the reparametrization, the open string
coupling is computed as
\begin{align}
 G_s &= \sqrt{\frac{\alpha \beta }{\cosh ^2 \theta \sin ^2 {\phi}
    (\beta + \tanh^2 \theta \cot^2 {\phi} \alpha)}}.
\end{align}
We further change the coordinates as
\begin{align}
 \sinh\theta &\to \sqrt{E^2 - 1} \sin\theta ,
 &\sin{\phi} &\to \sqrt{1 - C^2} \sin{\phi} ,
\end{align}
where $0 \leq \theta \leq \pi/2$. In the new
parametrization, the open string metric and the open string
coupling become
\begin{align}
 ds^2 &= - {d\theta}^2 + {d\phi} ^2
  + \frac{\tan ^2 \theta}
  {1 + \tan ^2 \theta \cot^2 {\phi} } d\lambda^2_- ,\\
 G_s &= \frac{1}{\sqrt{(E^2-1)(1-C^2)
    (\cos ^2 \theta \sin ^2 {\phi}
   + \sin ^2 \theta \cos ^2 {\phi})}}.
\end{align}
Therefore the Laplacian is given by
\begin{align}
 \Delta = - \frac{1}{\sin \theta} \partial_{\theta}
            \sin \theta \partial_{\theta}
          + \frac{1}{\sin {\phi}} \partial_{{\phi}}
            \sin {\phi}\partial_{{\phi}}
          + (\cot ^2 \theta + \cot ^2 {\phi})
            \partial_{\lambda_-}^2 .
            \label{laplaceAA1}
\end{align}
The AA-type boundary conditions are given by $J^A = \bar J^A$ and $K^A
= \bar K^A$ before gauging, and $J^3 + \bar J^3$ and $K^3 + \bar
K^3$ generate transformation along $\lambda_-$. This is once again
consistent with the form of the Laplacian \eqref{laplaceAA1}.

\subsubsection{AB-brane}

Let us study the case with $\kappa > 0 , |\xi| = 1$ ($\cos \phi =
0$); namely the D1-brane. Now that the time coordinate in the bulk
is $\lambda_-$, we can use $(\lambda_-,\theta)$ as the
coordinate system of the worldvolume.
Then, the  DBI action of D1-brane is given by
\begin{align}
 S&=-\tau_1 \int d \lambda_- d \theta e^{-\Phi} \sqrt{-\det (g+F)}
   \nonumber \\
  &=-\tau_1 \int d \lambda_- d \theta
   \sqrt{\sinh ^2 \theta - \cosh ^2 \theta f^2}
\end{align}
with $F_{\lambda_- \theta}=f$. The Gauss constraint leads to
\begin{align}
 \Pi = \frac{\delta {\cal L}}{\delta f}
     = \frac{\tau_1 \cosh ^2 \theta f}
       {\sqrt{\sinh ^2 \theta - \cosh ^2 \theta f^2}},
\end{align}
or
\begin{align}
 f^2 = \frac{\sinh ^2 \theta}
  {\frac{\tau_1^2}{\Pi^2} \cosh ^4\theta + \cosh ^2\theta}.
\end{align}
If we set $\Pi^2/\tau_1^2 = - \kappa ^2$, then the condition $f^2
\geq 0$ is consistent with the geometry $\cosh \theta \leq \kappa$
deduced in the previous subsection. However, the DBI action
\begin{align}
 S = - \tau_1 \int d \lambda_- d \theta
       \sqrt{\frac{\sinh ^2 \theta \cosh ^2 \theta}
                  {\cosh ^2 \theta - \kappa ^2}}
\end{align}
becomes imaginary, so the AB-type D1-brane is unphysical due to
the presence of supercritical electric flux.

Next let us examine D3-brane with $\kappa > 0, |\xi| < 1$. We turn
on the following field strengths
\begin{align}
 F_{\phi \lambda_-} &=0,
&F_{\theta  \lambda_+} &=0,
&F_{\lambda_+ \lambda_-} & = -1 , \nonumber\\
 F_{\theta \phi} &= F (\theta, \phi),
&F_{\phi \lambda_+} &= F_+ (\theta, \phi), &F_{\theta
\lambda_-} &= F_- (\theta, \phi) .
\end{align}
With the above choice of the field strength, we have
\begin{align}
 g+B+F =
  \begin{pmatrix}
   - \bar D \coth ^2 \theta & D-1 (=-\bar D) & 0 & - F_+ \\
   1 - D ( = \bar D) & \bar D \tan ^2 \phi & - F_- & 0  \\
    0 & F_- & 1 & F \\
   F_+ & 0 & - F & 1
   \end{pmatrix},
\end{align}
where  $\bar D^{-1} = 1- \tan^2 \phi \coth ^2 \theta$. Hence the
DBI action is given by
\begin{align}
 S = - \tau_3 \int  d \lambda_+ d \lambda_- d \theta d \phi
       \sqrt{ - {\rm sgn} \bar D \sinh ^2 \theta \cos ^2 \phi I }
\end{align}
with
\begin{align}
 I = 1 - F_-^2 \coth ^2 \theta + F_+^2 \tan ^2 \phi
           + F^2 - 2 F_- F_+ F - \bar D^{-1} F_-^2 F_+^2 .
\end{align}
Using the solution $F=F_+F_-$ to the equation $\frac{\delta
{\cal L}}{\delta F} = 0$, the DBI action is written as
\begin{align}
 S = - \tau_3 \int d \lambda_+ d \lambda_- d \theta d \phi
       \sqrt{ - {\rm sgn} \bar D
             (\sinh ^2 \theta - F_-^2 \cosh ^2 \theta)
             (\cos ^2 \phi + F_+^2 \sin ^2 \phi)}.
\end{align}
The equations of motion for $A_{\lambda_+}$ and $A_{\lambda_-}$
are given by
\begin{align}
 \partial_{\theta}
  \left(\frac{\delta {\cal L}}{\delta F_-}\right) &= 0,
&\partial_{\phi}
  \left(\frac{\delta {\cal L}}{\delta F_+}\right) &= 0,
\end{align}
which lead to
\begin{align}
C_1 &= \frac{\cosh ^2 \theta F_-}
     {\sqrt{- {\rm sgn} \bar D
      (\sinh ^2 \theta  - F_-^2 \cosh ^2 \theta)}} ,
&C_2 &= \frac{\sin ^2 \phi F_+}
     {\sqrt{\cos ^2 \phi + F_+^2 \sin ^2 \phi}}
\end{align}
with constants $C_{1}$ and $C_{2}$. The above equations imply
\begin{align}
 F_-^2 &= \frac{\sinh ^2 \theta}
  {- \frac{{\rm sgn} \bar D}{C_1^2} \cosh ^4 \theta + \cosh ^2 \theta} ,
 &F_+^2 &= \frac{\cos ^2 \phi}
  {\frac{1}{C_2^2} \sin ^4 \phi - \sin ^2 \phi} .
\end{align}
If we set $C_1^2 = {\rm sgn} \bar D \kappa ^2$ and $C_2^2 =
\xi^2$, then the condition that $F_{\pm}^2$ be non-negative gives
consistent result of the geometry \eqref{D3-braneAB1}.
The DBI action is
\begin{align}
 S = - \tau_3 \int d \theta d\phi d \lambda_- d \lambda_+
 \sqrt{\frac{\sinh^2 \theta \cosh ^2 \theta \sin^2 \phi \cos ^2 \phi}
 {{\rm sgn} \bar D (\kappa ^2 - \cosh ^2 \theta)(\sin ^2 \phi -\xi^2)}} ,
\end{align}
which is real for $\bar D>0$ and imaginary for $\bar D<0$. Hence
the D-brane is physical for $\bar D>0$ and unphysical for $\bar
D<0$ due to the presence of supercritical electric fields, as in
the previous case. In this case, however, there are CTCs even on
the physical region, since $\lambda_+$, which is now time, runs along
the CTCs in the region.

Because there are CTCs on the brane, it is very important to see whether
the wave functions on the brane behave pathologically.
We assume that $\bar D>0$ and $C_1^2 > 0$.
Then, the open string metric and the open string coupling are
\begin{align}
 ds_{open}^2& = - \alpha {d\theta}^2 + \beta {d\phi} ^2
  + \beta \cot ^2 {\phi} (d \lambda_+ - F_- d \theta )^2
  + \alpha \tan ^2 \theta (d \lambda_- + F_+ d {\phi})^2 ,\\
  G_s &= \sqrt{\frac{\alpha \beta}{\cosh ^2 \theta \sin ^2 {\phi}}},
\end{align}
where
\begin{align}
 \alpha &= - 1 + \coth ^2 \theta F_-^2
         = \frac{\cosh ^2 \theta}{- \cosh ^2 \theta + C_1^2},
&\beta & = 1 + \tan ^2 {\phi} F_+^2
         = \frac{\sin ^2 {\phi}}{\sin ^2 {\phi} - C_2^2} .
\end{align}
Again the time coordinate is $\theta$ in the open string metric,
and it ranges from $0$ to $\arccosh |C_1|$. We can further
simplify the metric by making the following transformation
\begin{align}
   \lambda_+ - \int^\theta F_- (x) dx &\to \lambda_+ ,
   &\lambda_- + \int^{{\phi}} F_+(x) dx &\to \lambda_- ,
\end{align}
and
\begin{align}
 \sinh \theta &\to \sqrt{C_1^2 - 1} \sin \theta ,
 &\cos {\phi} &\to \sqrt{1-C_2^2} \cos {\phi} .
\end{align}
After changing the coordinates, the time $\theta$ runs $0 \leq
\theta \leq \pi/2$. The open string metric and the open string
coupling become
\begin{align}
 ds_{open}^2 &= - {d\theta}^2 + {d\phi} ^2
 + \cot ^2 {\phi}d ^2 \lambda_+ + \tan ^2 \theta d ^2 \lambda_- ,\\
 G_s &=\frac{1}{ \sqrt{(C_1^2 - 1)(1-C_2^2) \cos^2 \theta \sin^2 {\phi}}} .
\end{align}
Therefore the Laplacian is given by
\begin{align}
 \Delta = - \frac{1}{\sin \theta} \partial_{\theta}
                     \sin \theta \partial_{\theta}
          + \frac{1}{\cos {\phi}} \partial_{{\phi}}
                     \cos {\phi} \partial_{{\phi}}
          + \tan ^2 {\phi}\partial_{\lambda_+}^2
          + \cot ^2 \theta  \partial_{\lambda_-}^2 .
\end{align}
No harmful thing seems to happen because the eigenfunction
equation of the Laplacian can be reduced to harmonic analysis
for $SL(2,\br)/U(1)$ and $SU(2)/U(1)$ WZW models.
Thus, we may conclude that CTCs on the brane are not so
pathological in the same meaning as CTCs in the bulk.
In section \ref{WF} we make a more precise argument on this point.

\subsubsection{BA-brane}

The BA-brane is very similar to the AB-brane as we will see below.
D1-brane of this type is examined first. The DBI action is given by
\begin{align}
 S &= - \tau_1 \int  d \lambda_+ d \theta
   e^{-\Phi} \sqrt{- \det (g + F)}
  \nonumber \\
   &= - \tau_1 \int d \lambda_+ d \theta
    \sqrt{\cosh ^2 \theta - \sinh ^2 \theta f^2 },
\end{align}
where we set $F_{\lambda_+ \theta} = f$.
The Gauss constraint leads to
\begin{align}
 \frac{ \delta {\cal L}}{\delta f}
   = \frac{\tau_1 \sinh ^2 \theta f}
     {\sqrt{\cosh ^2 \theta - \sinh ^2 \theta f^2 }} = \Pi
\end{align}
with a constant $\Pi$, which can be written as
\begin{align}
 f^2 = \frac{\cosh^2 \theta}
 {\frac{\tau_1^2 }{\Pi^2}\sinh^4 \theta + \sinh^2 \theta} .
\end{align}
If we set $\Pi^2/\tau_1^2 = - \kappa^2$, then we can reproduce the
geometry $\sinh \theta \leq \kappa$. However, the DBI action
becomes
\begin{align}
 S = - \tau_1 \int d \lambda_+ d \theta
 \sqrt{\frac{\sinh ^2 \theta \cosh ^2 \theta}{\sinh ^2 \theta - \kappa
 ^2}},
\end{align}
which is imaginary for $\sinh \theta \leq \kappa$. Therefore the
D1-brane is unphysical due to the presence of supercritical
electric field.

Next, let us consider D3-brane with $\kappa > 0, |\xi| < 1$.
Exciting $F_{\theta \lambda_+} = F_+,F_{\phi \lambda_-} = F_-,
F_{\theta \phi} = F$, the DBI action is given by
\begin{align}
 S = - \tau_3 \int d \lambda_+ d \lambda_- d \theta d \phi
       \sqrt{{\rm sgn} D \cosh ^2 \theta \sin ^2 \phi I }
\end{align}
where
\begin{align}
 I = - 1 + (F_+)^2 \tanh ^2 \theta - (F_-)^2 \cot ^2 \phi
           - F^2 + 2 F_+ F_- F - D^{-1}F_+^2 F_-^2 .
\end{align}
Inserting the solution $F=F_+F_-$ to the equation
$0=\frac{\delta {\cal L}}{\delta F}$,
the DBI action becomes
\begin{align}
 S = - \tau_3 \int d \lambda_+ d \lambda_- d \theta d \phi
       \sqrt{-{\rm sgn}D (\cosh ^2 \theta - F_+^2 \sinh ^2 \theta)
             (\sin ^2 \phi + F_-^2 \cos ^2 \phi)}.
\end{align}
Since the equations of motion for $A_{\lambda_+}, A_{\lambda_-}$
are given by
\begin{align}
 \partial_{\theta}
  \left(\frac{\delta {\cal L}}{\delta F_+}\right) &= 0,
&\partial_{\phi}
  \left(\frac{\delta {\cal L}}{\delta F_-}\right) &= 0,
\end{align}
we obtain
\begin{align}
C_1 &= \frac{\sinh ^2 \theta F_+}
     {\sqrt{-{\rm sgn}D (\cosh ^2 \theta  - F_+^2 \sinh ^2 \theta )}} ,
&C_2 &= \frac{\cos ^2 \phi F_-}
     {\sqrt{\sin ^2 \phi + F_-^2 \cos ^2 \phi}}
\end{align}
with constants $C_{1}$ and $C_{2}$. Therefore we have
\begin{align}
 F_+^2 &= \frac{\cosh ^2 \theta}
  {-\frac{{\rm sgn}D}{C_1^2} \sinh ^4 \theta + \sinh ^2 \theta} ,
 &F_-^2 &= \frac{\sin ^2 \phi}
  {\frac{1}{C_2^2} \cos ^4 \phi - \cos ^2 \phi} .
\end{align}
Let us suppose $C_1^2 = {\rm sgn} D \kappa ^2$ and $C_2^2 = \xi^2
$. Then, we reproduce the condition $\sinh ^2 \theta \leq \kappa
^2 $ and $\cos ^2 \phi \geq \xi^2$ from the fact that $F_{\pm}^2$
should be non-negative. The DBI action
\begin{align}
 S = - \tau_3 \int d \theta d\phi d \lambda_- d \lambda_+
 \sqrt{\frac{\sinh^2 \theta \cosh ^2 \theta \sin^2 \phi \cos ^2 \phi}
 {{\rm sgn} D (\kappa ^2 - \sinh ^2 \theta)(\cos ^2 \phi -\xi^2)}}
\end{align}
is real for $D > 0$ and imaginary for $D < 0$. Therefore the
D-brane is physical for $D>0$ and unphysical for $D<0$. CTCs also
exit on the branes in this case.

The Laplacian computed using the open string metric is very
similar to the AB-type case. We again assume that $D>0$ and $C_1^2 >
0$. The open string metric and the open string coupling are then
given by
\begin{align}
 ds^2_{open}& = - \alpha {d\theta}^2 + \beta {d\phi} ^2
  + \alpha \coth ^2 \theta  (d \lambda_+ + F_- d {\phi})^2
  + \beta \tan ^2 {\phi}(d \lambda_- - F_+ d \theta )^2 ,\\
  G_s &= \sqrt{\frac{\alpha \beta}{\sinh ^2 \theta \cos ^2 {\phi}}},
\end{align}
where
\begin{align}
 \alpha &= - 1 + \tanh ^2 \theta F_+^2
         = \frac{\sinh ^2 \theta}{- \sinh ^2 \theta + C_1^2},
&\beta & = 1 + \cot ^2 {\phi} F_-^2
         = \frac{\cos ^2 {\phi}}{\cos ^2 {\phi} - C_2^2} .
\end{align}
We rewrite
\begin{align}
    \lambda_+ + \int^{{\phi}} F_- (x) dx &\to \lambda_+,
   &\lambda_- - \int^{\theta} F_+(x) dx &\to \lambda_-,
\end{align}
and change
\begin{align}
 \cosh \theta &\to \sqrt{1+C_1^2} \cos \theta ,
 &\sin {\phi} &\to \sqrt{1-C_2^2} \sin {\phi} .
\end{align}
The time direction in the new coordinate $\theta$ runs $0 \leq
\theta \leq \pi/2$. In the new coordinate system, the open string
metric and the open string coupling are
\begin{align}
 ds^2_{open} &= - {d\theta}^2 + {d\phi} ^2
 + \cot ^2 \theta d \lambda_+^2 + \tan ^2 {\phi} d \lambda_-^2 ,\\
 G_s &=\frac{1}{ \sqrt{(1+C_1^2)(1-C_2^2) \sin^2 \theta \cos^2 {\phi}}}.
\end{align}
The Laplacian is given by
\begin{align}
 \Delta = - \frac{1}{\cos \theta} \partial_{\theta}
                     \cos \theta \partial_{\theta}
          + \frac{1}{\sin {\phi}} \partial_{{\phi}}
                     \sin {\phi} \partial_{{\phi}}
          + \tan ^2 \theta \partial_{\lambda_+}^2
          + \cot ^2 {\phi}\partial_{\lambda_-}^2 .
\end{align}
Since the AB-brane is replaced by the BA-brane, roughly speaking, the
roles of $SU(2)$ and $SL(2,\br)$ parts are exchanged.

\subsubsection{BB-brane}

Again the BB-branes are very similar to the AA-branes.
Since the D0-brane corresponds to the case with $\sin \phi = 1$,
the DBI action becomes
\begin{align}
 S &= - \tau_0 \int d \lambda_- e^{-\Phi} \sqrt{- \det g} \nonumber \\
   &= - \tau_0 \int d \lambda_-
    \sqrt{\sinh ^2 \theta - \cosh ^2 \theta  (\dot \theta )^2 }.
\end{align}
This system has a symmetry under time-translation, thus the energy
is conserved. The total energy is given by
\begin{align}
 E = \frac{\delta {\cal L}}{\delta \dot \theta} \dot \theta - {\cal L}
  = \frac{\tau_0 \sinh ^2 \theta}
         {\sqrt{\sinh^2 \theta - \cosh ^2 \theta (\dot \theta )^2}},
\end{align}
which  can be represented by
\begin{align}
 (\dot \theta )^2 = \frac
 {- \frac{\tau^2 _0}{E^2}
  \sinh ^4 \theta + \sinh ^2\theta }{\cosh ^2 \theta} .
\end{align}
This equation implies the motion of D0-brane
\begin{align}
 \sinh (\lambda_ - - \lambda_-^0) \sin \theta = \frac{E}{\tau _0 } ,
\end{align}
which reproduces the results obtained from the group theoretical considerations.
The DBI action
\begin{align}
 S = - \tau_0 \int d \lambda_- \frac{\tau_0 \sinh ^2 \theta}{E}
\end{align}
is real, so the D0-brane is physical.

For the D2-brane, we excite only $F_{\lambda_+ \phi} = F$. The
DBI action can be constructed from
\begin{align}
-\det (g+B+F) = - \bar D(1+{(\theta  ')}^2)
 + \bar D \coth ^2 \theta (\dot \theta)^2
-2\bar DF\dot \theta \theta ' - F^2 (\bar D\tan ^2 \phi +
(\dot\theta)^2),
\end{align}
where we use $\dot {\theta} = \frac{\partial \theta}{\partial
\lambda_-}$.
Assuming that $F=- \theta'/\dot \theta$, the equation of motion
for $F$ ($\partial_{\phi} \left(\frac{\delta \cal L}{\delta F}
\right) =0$) is reduced to
\begin{align}
 \frac{\frac{\theta '}{\dot \theta} \sin ^2 \phi}
      {\sqrt{\cos ^2 \phi + \sin ^2 \phi
   \frac{{(\theta ')}^2}{(\dot \theta)^2}}} = C ,
\end{align}
or
\begin{align}
 \frac{{(\theta')}^2}{(\dot \theta)^2}
  = \frac{\cos ^2 \phi}
    {\frac{1}{C^2} \sin ^4 \phi - \sin ^2 \phi }
    \label{diffbb1}
\end{align}
with a constant $C$. Since the energy momentum tensor is given by
\begin{align}
 T^{\lambda_-}_{~ \lambda_-} &=
 - \tau_2 {\rm sgn} \bar D  \sinh ^2 \theta
   \sqrt{\frac{\cos ^2 \phi + \sin ^2 \phi
   \frac{{(\theta ')}^2}{(\dot \theta)^2}}{
   - {\rm sgn} \bar D (\sinh ^2 \theta
    - \cosh ^2 \theta (\dot \theta)^2)}},
 & T^{\phi}_{~ \lambda_-} & = 0 ,
\end{align}
the conservation of the energy momentum tensor
\begin{align}
 \partial_{\lambda_-} T^{\lambda_-}_{ ~ \lambda_-}
  + \partial_{\phi} T^{\phi}_{ ~ \lambda_-} = 0
\end{align}
 leads to
\begin{align}
 E = \frac{\sinh ^2 \theta}{\sqrt{- {\rm sgn} \bar D
       (\sinh ^2 \theta - \cosh ^2 \theta (\dot \theta)^2)}},
\end{align}
or
\begin{align}
 (\dot \theta)^2 =
 \frac{\frac{{\rm sgn} \bar D}{E^2} \sinh ^4 \theta + \sinh ^2 \theta}
      {\cosh ^2 \theta} .
      \label{diffbb2}
\end{align}
If we use $E^2 = - {\rm sgn} D \kappa ^2 $ and $ C^2 = \xi ^2$,
then we reproduce the geometry \eqref{D2'-brane1} from equations
\eqref{diffbb1} and \eqref{diffbb2}. The DBI action now becomes
\begin{align}
 S = - \tau_2 \int d \lambda_+ d \lambda_- d \phi
  \sqrt{\frac{\sinh ^4 \theta
              \sin ^2 \phi \cos ^2 \phi}
             { - {\rm sgn} \bar D \kappa ^2 (\sin ^2 \phi - \xi ^2 ) }},
\end{align}
which is real for $\bar D < 0$ and imaginary for $\bar D > 0$.
Therefore the D-brane is physical for $\bar D < 0$ and unphysical
for $\bar D > 0$ due to the tachyonic behavior. The time direction
for $\bar D < 0$ is $\lambda_-$, so there are no CTCs on the
brane.

Wave functions on the D-branes seem to be also similar to the AA-type
case.
Suppose that $\bar D < 0$, then the open string metric is computed as
\begin{align}
 ds^2_{open} &= \beta {d\phi} ^2 -
 \alpha (\dot \theta d \lambda_- + \theta ' d {\phi})^2
  + \frac{\alpha \beta \cot ^2 {\phi}}
  {\alpha + \beta \tanh ^2 \theta \cot^2 {\phi} }
   d\lambda^2_+
\end{align}
with
\begin{align}
\alpha &= - 1 + \tanh ^2 \theta \frac{1}{(\dot \theta)^2}
         = \frac{\sinh ^2 \theta}{E^2 - \sinh ^2 \theta},
&\beta & = 1 + \tan ^2 {\phi}\frac{{(\theta ')}^2}{(\dot
\theta)^2}
         = \frac{\sin ^2 {\phi}}{\sin ^2 {\phi} - C^2} .
\end{align}
Changing the coordinate system from $(\lambda_+,\lambda_-, \phi)$
into $(\lambda_+, \theta, \phi)$, the open string metric and
the open string coupling can be rewritten as
\begin{align}
 ds^2_{open} &= - \alpha {d\theta}^2 + \beta {d\phi} ^2
  + \frac{\alpha \beta \cot ^2 \phi}
  {\alpha + \beta \tanh ^2 \theta \cot^2 {\phi} }
   d\lambda^2_+ ,\\
 G_s &= \sqrt{\frac{\alpha \beta }{\cosh ^2 \theta \sin ^2 {\phi}
    (\alpha + \tanh^2 \theta \cot^2 {\phi} \beta)}}.
\end{align}
We further change the coordinates as
\begin{align}
 \cosh\theta &\to \sqrt{E^2 + 1} \cos\theta ,
 &\cos {\phi} &\to \sqrt{1 - C^2} \cos{\phi} .
\end{align}
In the new parametrization ($0 \leq \theta \leq \pi/2$), the open
string metric and the open string coupling become
\begin{align}
 ds^2_{open} &= - {d\theta}^2 + {d\phi} ^2
  + \frac{\cot ^2 {\phi}}
  {1 + \tan ^2 \theta \cot^2 {\phi} } d\lambda^2_+ ,\\
 G_s &= \frac{1}{\sqrt{(E^2 + 1)(1-C^2)
    (\cos ^2 \theta \sin ^2 {\phi}
   + \sin ^2 \theta \cos ^2 {\phi})}}.
\end{align}
Therefore the Laplacian is
\begin{align}
 \Delta = - \frac{1}{\cos \theta} \partial_{\theta}
            \cos \theta \partial_{\theta}
          + \frac{1}{\cos {\phi}} \partial_{{\phi}}
            \cos {\phi}\partial_{{\phi}}
          + (\tan ^2 \theta + \tan ^2 {\phi})
            \partial_{\lambda_+}^2 .
\end{align}
Now that $J^3 - \bar J^2$ and $K^3 - \bar K^3$ generate parts of symmetry
left over on the brane, the Laplacian include
$\partial_{\lambda_+}$ both in $SU(2)$ and $SL(2,\br)$ sectors.

\subsection{Effective actions on D-branes in region $2'$}
\label{DBIregion2}

As we saw in subsection \ref{conjugacyw}, the D-branes in region $2'$
are qualitatively different from those in region 1. Fortunately,
the background in region $2'$ is just the one with replacing
$\lambda_+$ by $\lambda_-$ in region 1 if we use the region $- \pi/2 \leq \phi
\leq 0$ for the $SU(2)$ part. Therefore, we can analyze the DBI action
using the results in the previous subsection.
Despite of the similarity, we will see that wave functions on D-branes
in region $2'$ are quite different from those in region 1.

\subsubsection{AA-brane}

Now that we replace $\lambda_+$ by $\lambda_-$, the
BB-brane of region-1 with the replacement $\lambda_+ \leftrightarrow \lambda_-$ corresponds to
the AA-brane in this case. Thus, for the D2-brane, the non-trivial
components of gauge flux are
\begin{align}
 F_{\lambda_- \phi} &= -\frac{\theta '}{\dot \theta} ,
&F_{\lambda_- \lambda_+} &= -1 ,
&\frac{\partial \theta}{\partial \lambda_+} &= \dot \theta ,
&\frac{\partial \theta}{\partial \phi} &= \theta ' ,
\end{align}
and the equations of motion reduce to
\begin{align}
 \frac{{(\theta ')}^2}{(\dot \theta)^2}
  &= \frac{\cos ^2 \phi}
    {\frac{1}{C^2} \sin ^4 \phi - \sin ^2 \phi } ,
 &(\dot \theta)^2 &=
 \frac{\frac{{\rm sgn} \bar D}{E^2} \sinh ^4 \theta + \sinh ^2 \theta}
      {\cosh ^2 \theta} .
\end{align}
If we assume that $E^2 = {\rm sgn} D \kappa ^2 $ and $C^2 = \xi^2$,
then we reproduce the geometry \eqref{D2-brane1} computed from the group theory.
The DBI action
\begin{align}
 S = - \tau_2 \int d \lambda_+ d \lambda_- d \phi
  \sqrt{\frac{\sinh ^4 \theta
              \sin ^2 \phi \cos ^2 \phi}
             {{\rm sgn} \bar D \kappa ^2 (\sin ^2 \phi - \xi ^2 ) }}
\end{align}
is real for $\bar D > 0$ and imaginary for $\bar D < 0$. Therefore
the D-brane is physical for $\bar D > 0$ and is unphysical for
$\bar D < 0$ due to its tachyonic behavior. The time direction for
$\bar D > 0$ is $\lambda_-$, so there are CTCs on the brane
contrary to the BB-brane in region 1.

Suppose $\bar D > 0$,
then the open string metric and the open string coupling can be computed as
\begin{align}
 ds_{open}^2 &= \alpha d \theta ^2 + \beta d \phi ^2 -
 \frac{\alpha \beta \cot ^2 \phi}
 {\beta \tanh^2 \theta \cot^2 \phi - \alpha} d \lambda_- ^2, \\
 G_s &= \sqrt{\frac{\alpha \beta}
 {\cosh^2 \theta \sin ^2 \phi
 (\beta \tanh ^2 \theta \cot ^2 \phi - \alpha)}} ,
\end{align}
where
\begin{align}
 \alpha &= 1 - \tanh^2 \theta \frac{1}{(\dot \theta)^2}
        = \frac{\sinh^2 \theta}{\sinh^2 \theta + E^2} ,&
 \beta &= 1 + \tan ^2 \phi \frac{{(\theta ')}^2}{(\dot \theta)^2}
        = \frac{\sin ^2 \phi}{\sin ^2 \phi - C^2}.
\end{align}
As we saw in the previous sections, the open string metric can
take simpler form by performing the transformation of coordinates.
In this case, it is convenient to divide into three cases; 
$E^2 < 1$, $E^2 = 1$ and $E^2 > 1$. 
In the $SL(2,\br)$ WZW model, these cases
correspond to $dS_2$, light-cone and $H_2$ branes as mentioned
before, and hence we have to treat all these D-branes separately.
Since our model is a coset made from $SL(2,\br) \times SU(2)$ WZW
model, these three cases lead to different types of wave functions
on D-branes.

For $E^2 < 1$, we choose the transformations
\begin{align}
 \cosh \theta &\to \sqrt{1 - E^2} \cosh \theta ,
&\cos \phi&\to \sqrt{1 - C^2} \cos \phi ,
\end{align}
then the open string metric and the open string coupling become
\begin{align}
 ds_{open} ^2 &= d \theta ^2 + d \phi ^2
 - \frac{\cot ^2 \phi}{\tanh ^2 \theta \cot ^2 \phi - 1}
  d \lambda_- ^2 , \\
 G_s &= \frac{1}{\sqrt{(1-E^2)(1-C^2)
  (\sinh ^2 \theta \cos^2 \phi -
   \cosh^2 \theta \sin^2 \phi )}} .
\end{align}
Thus the Laplacian is given by
\begin{align}
 \Delta = \frac{1}{\cosh \theta} \partial_{\theta}
          \cosh \theta \partial_{\theta}
        + \frac{1}{\cos \phi} \partial_{\phi}
          \cos \phi \partial_{\phi}
        - ( \tanh ^2 \theta - \tan ^2 \phi ) \partial^2_{\lambda_-} .
\end{align}
For $E^2 > 1$, we change
\begin{align}
 \cosh \theta &\to \sqrt{E^2-1} \sinh \theta ,
&\cos \phi&\to \sqrt{1 - C^2} \cos \phi ,
\end{align}
where $\theta$ exists only when $\sinh \theta \geq 1/\sqrt{E^2 -1}$.
The open string metric and the open string coupling become
\begin{align}
 ds_{open} ^2 &= d \theta ^2 + d \phi ^2
 - \frac{\cot ^2 \phi}{\coth ^2 \theta \cot ^2 \phi - 1}
  d \lambda_- ^2, \\
 G_s &= \frac{1}{\sqrt{(E^2-1)(1-C^2)
  (\cosh ^2 \theta \cos ^2 \phi -
   \sinh^2 \theta \sin ^2 \phi )}} ,
\end{align}
thus the Laplacian operator is given by
\begin{align}
 \Delta = \frac{1}{\sinh \theta} \partial_{\theta}
          \sinh \theta \partial_{\theta}
        + \frac{1}{\cos \phi} \partial_{\phi}
          \cos \phi \partial_{\phi}
        - ( \coth ^2 \theta - \tan ^2 \phi ) \partial^2_{\lambda_-} .
\end{align}
In  both cases, the metrics and the Laplacians operators are very
similar to those in the bulk of whisker regions, which is
related to the fact that the D-branes are extended all the way to $\theta =
\infty$.

\subsubsection{AB-brane}

Here we can use the case of BA-brane of region 1
in order to construct AB-type D3-brane.
The non-trivial components of gauge flux can be read as
\begin{align}
 F_{\theta \lambda_-} &= F_+ ,
&F_{\phi \lambda_+} &= F_- ,
&F_{\theta \phi} &= F_+ F_- ,
\end{align}
and the solutions to the Gauss constraints are given by
\begin{align}
 F_+^2 &= \frac{\cosh ^2 \theta}
  {\frac{- {\rm sgn} D}{C_1^2} \sinh ^4 \theta + \sinh ^2 \theta} ,
 &F_-^2 &= \frac{\sin ^2 \phi}
    {\frac{1}{C_2^2} \cos ^4 \phi - \cos ^2 \phi } .
\end{align}
If we set $C_1^2 = - {\rm sgn} D \kappa ^2$ and $C_2^2 = \xi ^2$,
then $F_-^2 \geq 0$ leads to $\cos ^2 \phi \geq \xi^2$, which has
been obtained previously. The DBI action is
\begin{align}
 S = - \tau_3 \int d \theta d\phi d \lambda_- d \lambda_+
 \sqrt{\frac{\sinh^2 \theta \cosh ^2 \theta \sin^2 \phi \cos ^2 \phi}
 {-{\rm sgn} D (\sinh ^2 \theta + \kappa ^2)(\cos ^2 \phi -\xi^2)}} .
\end{align}
For $D < 0$ it is real so the D-brane is physical,
and for $D > 0$ it is imaginary so the D-brane is
unphysical. CTCs exist on the branes in this case as well.

Assuming $ D < 0$,
the open string metric and the open string coupling are
\begin{align}
 ds_{open}^2 &= \alpha d \theta ^2 + \beta d \phi ^2
       - \alpha \coth ^2 \theta d \lambda_- ^2
       + \beta \tan ^2 \phi d \lambda_+ ^2, \\
 G_s &= \sqrt{\frac{\alpha \beta}{\sinh^2 \theta \cos ^2 \phi}} ,
\end{align}
where
\begin{align}
 \alpha &= 1 - \tanh^2 \theta F_+^2
        = \frac{\sinh^2 \theta}{\sinh^2 \theta + C_1^2} ,&
 \beta &= 1 + \cot ^2 \phi F_-^2
        = \frac{\cos ^2 \phi}{\cos ^2 \phi - C_2^2}.
\end{align}
We consider two cases $C^2_1 < 1$ and $C_1^2 > 1$ also in this case.
For $C_1^2 <1$, we change the coordinates as
\begin{align}
 \cosh \theta &\to \sqrt{1 - C_1^2} \cosh \theta ,
&\sin \phi&\to \sqrt{1 - C_2^2} \sin \phi ,
\end{align}
then the open string metric and the open string coupling become
\begin{align}
 ds_{open} ^2 &= d \theta ^2 + d \phi ^2
  - \coth ^2 \theta d \lambda_- ^2 + \tan ^2 \phi d \lambda_+ ^2, \\
 G_s &= \frac{1}
 {\sqrt{(1-C_1^2)(1-C_2^2) \sinh ^2 \theta \cos^2 \phi}}.
\end{align}
The Laplacian operator is given by
\begin{align}
 \Delta = \frac{1}{\cosh \theta} \partial_{\theta}
          \cosh \theta \partial_{\theta}
        + \frac{1}{\sin \phi} \partial_{\phi}
          \sin \phi \partial_{\phi}
        - \tanh ^2 \theta \partial^2_{\lambda_-}
        + \cot ^2 \phi \partial^2_{\lambda_+} .
\end{align}
For $C_1^2 > 1$, we change
\begin{align}
 \cosh \theta &\to \sqrt{C_1^2-1} \sinh \theta ,
&\sin \phi&\to \sqrt{1 - C_2^2} \sin \phi ,
\end{align}
then the open string metric and the open string coupling become
\begin{align}
 ds_{open} ^2 &= d \theta ^2 + d \phi ^2
    - \tanh ^2 \theta d \lambda_- ^2
   + \tan ^2 \phi d \lambda_+ ^2, \\
 G_s &= \frac{1}{\sqrt{(C_1^2-1)(1-C_2^2)
  \cosh ^2 \theta \cos ^2 \phi }}.
\end{align}
The Laplacian is
\begin{align}
 \Delta = \frac{1}{\sinh \theta} \partial_{\theta}
          \sinh \theta \partial_{\theta}
        + \frac{1}{\sin \phi} \partial_{\phi}
          \sin \phi \partial_{\phi}
        -  \coth ^2 \theta \partial^2_{\lambda_- }
        + \cot ^2 \phi \partial^2_{\lambda_+} .
\end{align}
It might be interesting to notice that for $SL(2,\br)$ part of the above
two cases, the open string metrics or the Laplacian operators are of the
form T-dual to each other along $\lambda_-$.

\subsubsection{BA-brane}
To study the BA-brane one can use the result of AB-brane of region
1. Then the non-trivial components of gauge flux are
\begin{align}
 F_{\lambda_- \lambda_+} &= -1,
&F_{\theta \lambda_-} &= F_+ ,
&F_{\phi \lambda_+} &= F_- ,
&F_{\theta \phi} &=F_+ F_- ,
\end{align}
and the solutions to the Gauss constraints are
\begin{align}
 F_-^2 &= \frac{\sinh ^2 \theta}
  {\frac{- {\rm sgn} \bar D}{C_1^2} \cosh ^4 \theta + \cosh ^2 \theta} ,
 &F_+^2 &= \frac{\cos ^2 \phi}
    {\frac{1}{C_2^2} \sin ^4 \phi - \sin ^2 \phi } .
\end{align}
If we assign $C_1^2 = -{\rm sgn} \bar D \kappa ^2$ and $C_2^2 =
\xi^2$, then $F_{+}^2 \geq 0$ means $\sin ^2 \phi \geq \xi^2$.
Since the DBI action is
\begin{align}
 S = - \tau_3 \int d \theta d\phi d \lambda_- d \lambda_+
 \sqrt{\frac{\sinh^2 \theta \cosh ^2 \theta \sin^2 \phi \cos ^2 \phi}
 {-{\rm sgn} \bar D (\cosh ^2 \theta + \kappa)(\sin ^2 \phi -\xi^2)}} ,
\end{align}
the D-brane is physical for $\bar D < 0$ and unphysical for $\bar
D > 0$. There are CTCs on the D-brane as well.

Assuming $ \bar D < 0$,
the open string metric and the open string coupling are
\begin{align}
 ds_{open}^2 &= \alpha d \theta ^2 + \beta d \phi ^2
       + \beta \cot ^2 \phi d \lambda_- ^2
       - \alpha \tanh ^2 \theta d \lambda_+ ^2, \\
 G_s &= \sqrt{\frac{\alpha \beta}{\cosh^2 \theta \sin ^2 \phi}} ,
\end{align}
where
\begin{align}
 \alpha &= 1 - \coth^2 \theta F_-^2
        = \frac{\cosh^2 \theta}{\cosh^2 \theta + C_1^2} ,&
 \beta &= 1 + \tan ^2 \phi F_+^2
        = \frac{\sin ^2 \phi}{\sin ^2 \phi - C^2_2}.
\end{align}
Because the B-brane in $SL(2,\br)$ WZW model is $AdS_2$ brane in
$AdS_3$, we do not need to divide into different cases. Changing
the coordinates as
\begin{align}
 \sinh \theta &\to \sqrt{1 + C_1^2} \sinh \theta ,
&\cos \phi&\to \sqrt{1 - C_2^2} \cos \phi ,
 \label{BAnewcoord}
\end{align}
the open string metric and the open string coupling become
\begin{align}
 ds_{open} ^2 &= d \theta ^2 + d \phi ^2
  + \cot ^2 \phi d \lambda_- ^2 - \tanh ^2 \theta d \lambda_+ ^2, \\
 G_s &= \frac{1}
 {\sqrt{(1+C_1^2)(1-C_2^2) \cosh ^2 \theta \sin^2 \phi}}.
\end{align}
The Laplacian operator is
\begin{align}
 \Delta = \frac{1}{\sinh \theta} \partial_{\theta}
          \sinh \theta \partial_{\theta}
        + \frac{1}{\cos \phi} \partial_{\phi}
          \cos \phi \partial_{\phi}
        + \tan ^2 \phi\partial^2_{\lambda_-}
        - \coth ^2 \theta \partial^2_{\lambda_+} .
\end{align}

\subsubsection{BB-brane}

Here, once again, we should use the results of AA-brane of region 1 with $\lambda_+
\leftrightarrow \lambda_-$. The non-trivial components of gauge
flux are
\begin{align}
 F_{\lambda_+ \phi} & = -\frac{ \theta '}{\dot \theta} ,
&\frac{\partial \theta}{\partial \lambda_+} &= \dot \theta ,
&\frac{\partial \theta}{\partial \phi} &= \theta ' ,
\end{align}
and the equations of motion reduce to
\begin{align}
 \frac{{(\theta ')}^2}{(\dot \theta)^2}
  &= \frac{\sin ^2 \phi}
    {\frac{1}{C^2} \cos ^4 \phi - \cos ^2 \phi } ,
 &(\dot \theta)^2 &=
 \frac{\frac{{\rm sgn} D}{E^2} \cosh ^4 \theta + \cosh ^2 \theta}
      {\sinh ^2 \theta} .
\end{align}
If we assign $C^2 = \xi^2$ and $E^2 = {\rm sgn} D \kappa ^2$,
then we can reproduce the geometry \eqref{D2'-brane2} computed from the
group theory. The DBI action in this case
\begin{align}
 S = - \tau_2 \int d \lambda_+ d \lambda_- d \phi
  \sqrt{\frac{\cosh ^4 \theta
              \sin ^2 \phi \cos ^2 \phi}
             { {\rm sgn} D \kappa ^2 (\cos ^2 \phi - \xi ^2 ) }}
\end{align}
is real for $D > 0$ and imaginary for $D < 0$. Thus the D-brane is
physical for $D > 0$ and unphysical for $D < 0$. For $D > 0$,
$\lambda_+$ becomes time-coordinate, so CTC also exits on the D-brane.

Assuming $D > 0$, the
open string metric and the open string coupling are
\begin{align}
 ds_{open}^2 &= \alpha d \theta ^2 + \beta d \phi ^2 -
 \frac{\alpha \beta \tanh ^2 \theta}
 {\beta - \alpha \tanh^2 \theta \cot^2 \phi } d \lambda_+ ^2, \\
 G_s &= \sqrt{\frac{\alpha \beta}
 {\cosh^2 \theta \sin ^2 \phi
 (\beta - \alpha \tanh ^2 \theta \cot ^2 \phi )}} ,
\end{align}
where
\begin{align}
 \alpha &= 1 - \coth^2 \theta \frac{1}{(\dot \theta)^2}
        = \frac{\cosh^2 \theta}{\cosh^2 \theta + E^2} , &
 \beta &= 1 + \cos ^2 \phi \frac{{(\theta ')}^2}{(\dot \theta)^2}
        = \frac{\cos ^2 \phi}{\cos ^2 \phi - C^2}.
\end{align}
Changing the variables
\begin{align}
 \sinh \theta &\to \sqrt{1 + E^2} \sinh \theta ,
&\sin \phi&\to \sqrt{1 - C^2} \sin \phi ,
\end{align}
the open string metric and the open string coupling become
\begin{align}
 ds_{open} ^2 &= d \theta ^2 + d \phi ^2
 - \frac{\tanh ^2 \theta }{1 - \tanh ^2 \theta \cot ^2 \phi }
  d \lambda_+ ^2 , \\
 G_s &= \frac{1}{\sqrt{(1+ E^2)(1-C^2)
  ( \cosh^2 \theta \sin^2 \phi -
    \sinh ^2 \theta \cos^2 \phi )}}.
\end{align}
The Laplacian operator is given by
\begin{align}
 \Delta = \frac{1}{\sinh \theta} \partial_{\theta}
          \sinh \theta \partial_{\theta}
        + \frac{1}{\sin \phi} \partial_{\phi}
          \sin \phi \partial_{\phi}
        - ( \coth ^2 \theta - \cot ^2 \phi ) \partial^2_{\lambda_+} .
\end{align}

%%%%%%%%%%%%%%%%%%%%%%%%%%%%%%%%%%%%%%%%%%%%%%%%%%%%%%%%%%%%%%%%%%%%%%%%%

\section{Wave functions}
\label{WF}

In this section, we would like to examine solutions to the
eigenfunction equations of the Laplacians for both the bulk
and the D-brane cases.
Closed string spectrum at low energy can be examined by using the
following effective Lagrangian for a scalar field $\Psi$
(wave function);
\begin{align}
 {\cal L} = \sqrt{- g} e^{- 2 \Phi} g^{\mu \nu}
   \partial_{\mu} \Psi \partial_{\nu} \Psi .
\end{align}
The equation for small fluctuations reduces to the eigenvalue
equation of the Laplacians \eqref{LBII}, \eqref{LB1}. Since the
expressions of Laplacian are separated into $SU(2)$ and
$SL(2,\br)$ parts, we study the wave functions in the next two
subsections separately. Open string spectrum on D-branes at
low energy can be read off similarly from the effective Lagrangian for
a scalar field
\begin{align}
 {\cal L} = \sqrt{ - G} e^{- \Phi_o} G^{ab}
   \partial_a \Psi \partial_b \Psi ,
\end{align}
where one has to use the open string metric and the open string coupling \cite{SW}.
Thus, we study the eigenvalue equation of the Laplacian given in terms of the
open string quantities in order to read off the open string spectrum.

\subsection{$SU(2)$ part}

Let us first consider the closed string case.%
\footnote{Wave functions in $SU(2)$ part were also studied in \cite{MMS}
for both closed strings and open strings.}
The Laplacian for $SU(2)$ part is given by \eqref{LBII}, \eqref{LB1}
\begin{align}
 \Delta &=
   \frac{1}{\cos \phi \sin \phi} \partial_{\phi}
   (\cos \phi \sin \phi) \partial_{\phi} +
   \tan ^2 \phi \partial_{\lambda_+}^2 + \cot ^2 \phi
    \partial_{\lambda_-}^2 \nonumber \\
      & = 4 \partial_y y (1-y) \partial_y
  + \frac{y}{1-y} \partial^2_{\lambda_+}
  + \frac{1-y}{y} \partial^2_{\lambda_-} ,
   \label{LySU2}
\end{align}
where we make a change of variable $y=\sin ^2 \phi$
in the second equation.
We investigate the eigenvalue equation of the Laplacian
\begin{align}
(\Delta + 4 \lambda) \Psi  (y,\lambda_+,\lambda_-) = 0 ,
\end{align}
and a solution to the equation is given by
\begin{align}
 \Psi (y,\lambda_+,\lambda_-) = y^{\frac{|m|}{2}} (1-y)^{\frac{|n|}{2}}
   e^{im\lambda_+ + in\lambda_-} F(y)
    \label{wfSU21}
\end{align}
with a hypergeometric function
\begin{align}
 F(y)=F \left(\frac{|m| + |n|}{2} + l + 1, \frac{|m| + |n|}{2} - l ,
        1+|m| ; y \right) .
        \label{wfSU22}
\end{align}
Here we set $\lambda = l(l+1) - \frac{m^2}{4} - \frac{n^2}{4}$.
The wave function must be regular at $y=0,1$ such that the
norm
$\int_0^1 dy y^{\frac{|m|}{2}} (1-y)^{\frac{|n|}{2}} |F|^2$
should be finite (see \cite{MMS}). The above solution is chosen to be
finite at $y=0$.
In order to see the finiteness at $y=1$, it is convenient to use a formula
\begin{align}
&\lim_{z \to 1-0}\left[F(\alpha,\beta,\gamma;z)
  -\sum_{n=0}^{k} (-1)^n
 \frac{\Gamma (\alpha + \beta - \gamma -n)\Gamma(\gamma)}
      {\Gamma(\alpha)\Gamma(\beta) n!} \right. \nonumber \\
  &\qquad\qquad \left. \times
 \frac{\Gamma(\gamma - \alpha + n)\Gamma(\gamma - \beta + n)}
      {\Gamma(\gamma - \alpha) \Gamma (\gamma - \beta)}
      (1-z)^{n+\gamma -\alpha -\beta}
\right] =
 \frac{\Gamma (\gamma)\Gamma(\gamma - \alpha - \beta)}
      {\Gamma (\gamma - \alpha) \Gamma (\gamma - \beta)} ,
\end{align}
where ${\rm Re} (\alpha + \beta - \gamma) > 0$ and $k$ is the maximal
integer number less than ${\rm Re} (\alpha + \beta - \gamma)$.
Because the coefficients of the second term in
the bracket must be zero, we have to
set $\beta = \frac{|m| + |n|}{2} - l = -s$ with $s=0,1,2,\cdots$.
Satisfying this condition, we can show that the right hand side is finite.
Note that the condition implies $-l \leq m,n \leq l$.

For the open string case we have two types of Laplacian operators
\begin{align}
 \Delta &= \frac{1}{\sin \phi} \partial_{\phi}
          \sin \phi \partial_{\phi}
 + \cot ^2 \phi \partial_{\lambda_-} ,
 &\Delta &= \frac{1}{\cos \phi} \partial_{\phi}
          \cos \phi \partial_{\phi}
 + \tan ^2 \phi \partial_{\lambda_+} .
\end{align}
Since the second one can be analyzed by replacing
$y=\sin ^2 \phi \leftrightarrow 1-y = \cos ^2 \phi$,
we concentrate on the first one.%
\footnote{In the parafermion theory $SU(2)/U(1)$, the difference between
two cases is important and they correspond to A-brane and B-brane by
gauging $U(1)$ \cite{MMS}.}
We rewrite the Laplacian as
\begin{align}
 \Delta = 4 y(1-y) \partial_y^2 + (4 - 6y) \partial_y
 + \frac{1-y}{y} \partial_{\lambda_-}^2 ,
\end{align}
and use the following ansatz for the solution to the equation
$(\Delta + 4 \lambda) \Psi = 0$:
\begin{align}
 \Psi (y , \lambda_-) = y^{\frac{|m|}{2}} e^{im\lambda_-} F(y).
  \label{wfoSU21}
\end{align}
The solution finite at $y=0$ to the eigenvalue equation is
\begin{align}
 F=F \left(\frac{|m|-l}{2},\frac{|m|+l+1}{2},1+|m|,y \right) ,
  \label{wfoSU22}
\end{align}
where we set $\lambda=l(l+1)-m^2$.
As seen in the previous sections, the boundary is always at $y=1$,
and hence we have to assign boundary conditions to $F(y)$ at $y=1$.
The condition $F(y=1)=0$ reduces to $l=|m|+1+2s$ with $s=0,1,2,\cdots$,
while the condition $\partial_y F(y)|_{y=1} = 0$ reduces to
$l=|m|+2s$. At this stage, we do not know what kind of boundary condition
we should assign. For instance, for B-branes in the parafermion
theory, we have to use both types of boundary condition \cite{MMS}.

\subsection{$SL(2,\mathbb{R})$ part}

In the closed string case, we have two types of
Laplacian  \eqref{LBII}, \eqref{LB1}
\begin{align}
 \Delta &= -
   \frac{1}{\cos \theta \sin \theta} \partial_{\theta }
   (\cos \theta \sin \theta) \partial_{\theta} +
   \tan ^2 \theta \partial_{\lambda_+}^2 + \cot ^2 \theta
    \partial_{\lambda_-}^2 , \\
 \Delta &=
   \frac{1}{\cosh \theta \sinh \theta} \partial_{\theta }
   (\cosh \theta \sinh \theta) \partial_{\theta} -
   \tanh ^2 \theta \partial_{\lambda_+}^2 - \coth ^2 \theta
    \partial_{\lambda_-}^2 .
\end{align}
Rewriting $y=\sin ^2 \theta$ in the first one and
$y=-\sinh^2 \theta$ in the second one, both the Laplacians
reduce to
\begin{align}
 \Delta = - 4 \partial_y y (1-y) \partial_y
  + \frac{y}{1-y} \partial^2_{\lambda_+}
  + \frac{1-y}{y} \partial^2_{\lambda_-} .
\end{align}
Note that the range of $y$ is $0\leq y \leq 1$ in the first case
and $y \leq 0$ in the second case.
Comparing with the Laplacian for $SU(2)$ part \eqref{LySU2},
the eigenfunctions are easily obtained from the $SU(2)$ counterpart
\eqref{wfSU21} with \eqref{wfSU22}
by replacing $|m| \to \pm i m$ and $|n| \to \pm i n$ (and renaming
$j$ as $l$).
There are two independent solutions among the four,
and we pick up one linear combination by assigning boundary conditions.
For the $SU(2)$ part, the normalizability condition uniquely
determines the linear combination, as seen above. However, for the
$SL(2,\br)$ part, many linear combinations are allowed,
and we have to appropriately choose them according to the physics required.

Let us first consider the region $2'$, and start from the bulk case,
which was already examined closely in \cite{NW2}.
We use $j > -\frac12$
or $j= -\frac12 + is$ with $s \geq 0$
to construct (delta-functional) normalizable wave functions
\begin{align}
\Psi = e^{im\lambda_- + in\lambda_+}
  y^{i\frac{m}{2}} (1-y)^{i\frac{n}{2}} F (y) .
   \label{nwf}
\end{align}
For $j > -\frac12$, the wave function is uniquely determined by
normalizability condition as
\begin{align}
 F(y) = (-y)^{- \frac{i(m+n)}{2} - j - 1}
  F \left(\frac{i(m+n)}{2}+j+1, \frac{i(-m+n)}{2}+j+1, 2j+2;
   \frac1y  \right)
   \label{discretewf}
\end{align}
up to normalization. This wave function decays like
$\Psi \sim e^{-2(j+1)\theta}$
at $\theta \to \infty$ and corresponds to a bound state
localized around $\theta = 0$.
For $j= -\frac12 + is$, the normalizability does not specify
unique linear combinations.\footnote{See \cite{Hikida} for a
detailed investigation on wave functions in Lorentzian $AdS_3$.}
A natural choice may be \cite{DVV,NW2}
\begin{align}
 F (y) &= \frac
   {\Gamma (\frac{i(m+n)}{2} + j + 1)\Gamma (\frac{i(m - n)}{2} + j + 1)}
  {\Gamma(1+im)\Gamma(2j+1)} \nonumber \\
 &\qquad \qquad \qquad \times
 F\left(\frac{i(m+n)}{2} + j + 1, \frac{i(m+n)}{2} - j , 1 + im;y\right) ,
   \label{continuouswf}
\end{align}
whose asymptotic behavior at $y \to 0$ is $\Psi \sim y^{i\frac{m}{2}}$.
Namely, there is only out-going wave near $y \sim 0$.
If we take $y \to - \infty$ ($\theta \to \infty$) limit,
then the wave function behaves
\begin{align}
  \Psi \sim e^{-\theta} [ e^{2is\theta} + R (j,m,n) e^{-2is\theta} ] ,
\end{align}
where the reflection coefficient is
\begin{align}
 R(j,m,n) = \frac{\Gamma(-2j-1) \Gamma (\frac{i(m+n)}{2} +j+1)
 \Gamma(\frac{i(m - n)}{2} + j + 1)}{\Gamma (2j+1)
 \Gamma (\frac{i(m+n)}{2} - j)\Gamma (\frac{i(m - n)}{2} - j)}.
 \label{rc}
\end{align}
Thus the wave function can be interpreted as the linear
combination of in-coming and out-going waves with the reflection
coefficient $R(j,m,n)$ \eqref{rc} near $\theta \sim \infty$.\footnote{Here we
assumed that $\lambda_-$ is the time and $m > 0$ as in
\cite{NW2}. Other cases can be analyzed in a similar manner.}
As shown in \cite{NW2}, the reflection
coefficient can be reproduced from the CFT analysis as the two
point function. Notice that the absolute value
\begin{align}
 |R(j,m,n)|^2 = \frac{\cosh \pi (m-2s) + \cosh \pi n}
                      {\cosh \pi (m+2s) + \cosh \pi n}
\end{align}
is always less than one, thus there must be transition to the
other regions. The existence of CTC affects only the labels
$m$ and $n$ to be integer through the gauge identification
\eqref{gauge}. This is analogous to the trivial closed time-like
curve case, such as, $S^1$ compactification of the time, which can be
undone easily. In the simple example, the energy is also quantized
due to the compactification.
It might be also interesting to notice that the wave functions
do not show any pathological behavior on the top of the domain wall
 $\tanh ^2 \theta \cot ^2 \phi = 1$ \eqref{singular-surface}.
This is because
the wave functions are given by the direct product of $SU(2)$ and
$SL(2,\br)$ parts.

Wave functions for the open strings are quite similar to those
for the closed strings.
Let us focus on BA-brane (or BB-brane) in region $2'$.
The wave functions are obtained
by linear combinations of \eqref{wfoSU21} with \eqref{wfoSU22}
with replacing $|m|$ by $\pm i m$.
The wave functions are like \eqref{discretewf} for $j > -1/2$ and
\eqref{continuouswf} for $j = -1/2 + i s$.
The only difference is that the original coordinates are shifted as
\eqref{BAnewcoord}, thus the reflection coefficient \eqref{rc}
includes extra factor $(\cosh r)^{2j+1}$ with $C_1 = \sinh r$
in the original coordinate system.
The same factor, which is associated with the boundary condition,
appears in the case of Euclidean $AdS_3$ \cite{PST} or Euclidean
two dimensional black hole \cite{RS}.

Next we consider the bulk case in region $III$.\footnote{%
We can obtain results for region $III$ by replacing
$\lambda_+ \leftrightarrow \lambda_-$ in those for region $II$.}
In this case, the parameter $y$ takes the value $0 \leq y \leq 1$,
and a natural choice of wave functions are analytic continuation
of those in region 2' ($y < 0$) as suggested in \cite{NW2}.
Therefore, we use again the same wave functions \eqref{discretewf}
for $j > -1/2$ and \eqref{continuouswf} for $j= -1/2 + is$ but with
a different parameter region $0 \leq y \leq 1$.
Note that in this case there is no asymptotic region
$y \sim - \infty$ but big crunch singularity at $y = 1$.
Let us focus on $j = -1/2 + is$ case.
Now that $\theta$ is the time direction,
the wave function \eqref{continuouswf} can be regarded
as a negative frequency mode of expansion of scalar field
\begin{align}
 \Psi \sim a e^{- im\lambda_- - in \lambda_+} y^{- i \frac{m}{2}}
      + a^{\dagger} e^{im\lambda_- + in \lambda_+} y^{i \frac{m}{2}}
\end{align}
near the big bang singularity at $y=0$. Here we use $m > 0$ as before
and denote $a,a^{\dagger}$ as the creation and annihilation operators respectively.
This means that the in-state is given by many particle state,
and this is consistent with the fact that there is a transition
from the region $2'$ as seen above.

In order to see the properties of wave functions in region $III$,
we set $m= - |m|$ (opposite to the previous case) and $n= - |n|$
and change the normalization as
\begin{align}
 \Psi = a e^{i|m|\lambda_- + i|n|\lambda_+} u (y)
      + a^{\dagger} e^{- i|m|\lambda_- - i|n|\lambda_+} u^* (y),
      \label{cosmowf}
\end{align}
with
\begin{align}
 u_b (y) = \frac{y^{- i\frac{|m|}{2}}(1-y)^{- i\frac{|n|}{2}}}{\sqrt{|m|}}
  F\left(- \frac{i(|m|+|n|)}{2} + j + 1
  , - \frac{i(|m|+|n|)}{2} - j , 1 - i|m|;y \right) .
\end{align}
Moreover, we choose the vacuum state as the initial state at $y=0$.
However, near the big crunch singularity at $y = 1$,
it might be natural to expand
\begin{align}
  \Psi \sim a e^{i|m|\lambda_- + i|n| \lambda_+} (1-y)^{i \frac{|n|}{2}}
      + a^{\dagger} e^{-i|m|\lambda_- - i|n| \lambda_+}
        (1-y)^{-i \frac{|n|}{2}} ,
\end{align}
namely, we use the expansion \eqref{cosmowf} with
\begin{align}
 u_c (y)  = \frac{y^{- i\frac{|m|}{2}}(1-y)^{i\frac{|n|}{2}}}{\sqrt{|n|}}
  F\left(- \frac{i(|m|-|n|)}{2} + j + 1
  , - \frac{i(|m|-|n|)}{2} - j , 1 + i|n|;1-y \right) .
  \label{outvac}
\end{align}
This expression is obtained by Bogolubov transformation
of the former one
\begin{align}
 u_c (m,n) &= \alpha (m,n) u_b (m,n) + \beta (m,n) u_b^* (-m,-n) ,
  \nonumber \\
  u_b (m,n) &= \alpha ^* (m,n) u_c (m,n) - \beta (m,n) u_c^* (-m,-n) ,
\end{align}
where
\begin{align}
 \alpha (m,n) &= \sqrt{\frac{|m|}{|n|}}
   \frac{\Gamma (1+i|m|) \Gamma (i |n|)}
   {\Gamma ( \frac{i(|m|+|n|)}{2}+j+1)\Gamma ( \frac{i(|m|+|n|)}{2}-j)} ,
    \nonumber \\
 \beta (m,n) &= - \sqrt{\frac{|m|}{|n|}}
   \frac{\Gamma (1-i|m|) \Gamma (i |n|)}
   {\Gamma ( \frac{i(-|m|+|n|)}{2}+j+1)\Gamma ( \frac{i(-|m|+|n|)}{2}-j)} .
\end{align}
These coefficients satisfy $|\alpha|^2 - |\beta|^2 = 1$.
The out vacuum is determined by the mode expansion with
\eqref{outvac}. Therefore, the transition between the initial vacuum state
and the final many particle state is (see, e.g., \cite{BD,Strominger,GS})
\begin{align}
 \gamma ^ * (j,m,n) = - \frac{\beta (m,n)}{\alpha (m,n)^*}
 = \frac{\Gamma (i|n|) \Gamma (- \frac{i(|m|+|n|)}{2} +j+1)
 \Gamma(- \frac{i(|m| + |n|)}{2} - j )}{\Gamma (-i|n|)
 \Gamma (- \frac{i(|m| - |n|)}{2} + j + 1)
 \Gamma (- \frac{i(|m| - |n|)}{2} - j)} .
 \label{ta}
\end{align}
The absolute value is related to the particle creation rate
\begin{align}
 |\gamma (j,m,n)|^2 = \frac{\cosh 2 \pi s
                      + \cosh \pi (|m| - |n|)}
                     {\cosh 2 \pi s
                      + \cosh \pi (|m| + |n|)} ,
\end{align}
which is of order ${\cal O}(1)$ in general.
In the string theory context, there is Hagedorn tower of
closed strings at high energy, and hence we expect many closed
string emission with high energy, which may lead to large back reaction.%
\footnote{Similar results are obtained in the models
of analytic continuation from $SL(2,\br)/U(1)$ WZW model
\cite{HikiTaka,TT}.}
In the open string case, the situation is very similar to
the bulk case as before, and we have boundary condition
dependent factor like $|C_1|^{i|n|}$ in the transition
amplitude \eqref{ta}.
The transition function has been conjectured to correspond to a two point
function \cite{Strominger,GS}, and it is therefore important to compare
our result to that obtained by CFT analysis.

%%%%%%%%%%%%%%%%%%%%%%%%%%%%%%%%%%%%%%%%%%%%%%%%%%%%%%%%%%%%%%%%%%%%%
%%%%%%%%%%%%%%%%%%%%%%%%%%%%%%%%%%%%%%%%%%%%%%%%%%%%%%%%%%%%%%%%%%%%%
\section{Conclusion} \label{conclusion}

In this paper, we have investigated properties of D-branes in
the Nappi-Witten (NW) model \cite{NW}, a gauged WZW model
based on the coset $(SL(2, {\mathbb R}) \times SU(2))/(U(1)\times
U(1))$. This four dimensional model consists of
cosmological regions with big bang/big crunch singularities and
whisker regions with closed time-like curves (CTCs). Since the
model can be defined as a gauged WZW model, strings in the
background should be well-behaved as emphasized in \cite{NW2}.

D-branes in this coset theory can be obtained by descending from
maximally symmetric and symmetry breaking branes in $SL(2,\br)
\times SU(2)$ \cite{Sarkissian,QS}. We have constructed DBI
actions for these D-branes and have shown that their classical
configurations are consistent with the group theoretic results.
Typically, D-branes see the background metric differently from
strings, and the D-brane metrics can be read from the DBI actions.
In particular, we have shown that the D-brane metrics do
not include singularities associated with big bang/big crunch
points in the cosmological regions. One may expect, naively, that
D0-brane feels the singularities more severely than string
because D0-brane is a point-like object, but it is not the case, as
we have discussed. This may have originated from the fact that the model
discussed in the paper is constructed as a gauged WZW model,
and hence even D-branes do not see any pathologies
related to the singularities. We have, however, neglected the back
reaction of the D-brane probes, and it would be important to
include it in order to obtain more definite conclusion. The back
reaction of strings or D-branes to the background could be investigated
by computing higher point functions or higher order corrections, which
deserves further study. In fact, it has been conjectured in
\cite{Pioline2,Pioline3} that the condensation of strings in twisted
sectors resolves the singularities in Misner space.

Open string spectra on D-branes in the low energy limit can be
read off from the eigenfunctions of the Laplacians represented in
terms of open string metrics and open string couplings \cite{SW}.
Even though the DBI actions and their classical configurations
seem to be complicated, the eigenvalue equations themselves are
quite simple due to the coset construction. In particular, we can
separate the $SU(2)$ and $SL(2,\br)$ parts, and the results reduce
to harmonic analysis in $SU(2)/U(1)$ and $SL(2,\br)/U(1)$ theory.
Therefore, we do not expect any pathology even if
D-brane wraps CTC in a whisker region.
However, in order to obtain more rigorous results on
open string spectra,
we have to face the full CFT analysis.
In particular, we should construct boundary states for
D-branes studied in this paper, for example, by following the
general methods given in \cite{QS}.\footnote{We may have to use
boundary states in Lorentzian $AdS_3$ recently proposed in \cite{Israel}.}
To make full CFT analysis, it might be useful to investigate a
simpler case, namely, open strings
in Misner space \cite{misner-boundary}, where it seems easier to
understand the origin of pathologies from singularities.
The superstring extension is also important in
order to study the stability of both the background and the D-branes.

%%%%%%%%%%%%%%%%%%%%%%%%%%%%%%%%%%%%%%%%%%%%%%%%%%%%%%%%%%%%%%%%%%%%%
\subsection*{Acknowledgement}

We would like to thank Massimo~Bianchi, Lorenzo~Cornalba, Gianfranco~Pradisi,
Soo-Jong~Rey, Augusto~Sagnotti, Yassen~Stanev and Yuji~Sugawara
for encouragement and useful discussions. YH is grateful
to KEK for the hospitality, where a part of this work was done.
RRN and KLP would like to acknowledge the hospitality of
the Abdus Salam ICTP where a part of this work was done.
The work of YH was supported by Brain Korea 21 Project. The work of RRN and KLP was
supported in part by INFN, by the MIUR-COFIN contract 2003-023852,
by the EU contracts MRTN-CT-2004-503369 and MRTN-CT-2004-512194,
by the INTAS contract 03-51-6346, and by the NATO grant
PST.CLG.978785.

%%%%%%%%%%%%%%%%%%%%%%%%%%%%%%%%%%%%%%%%%%%%%%%%%%%%%%%%%%%%%%%%%%%%%

\appendix

\section{D-branes in Misner space}
\label{Misner}

In this appendix, we would like to see some
properties of D-branes in the Misner space. Misner space
\cite{Misner} can be obtained by orbifolding a two dimensional
Minkowski space-time by a Lorentz boost $x^{\pm} = e^{\pm 2\pi}
x^{\pm}$. The space-time can be divided into four regions across
the line $x^+ x^- = 0$. Two regions $x^+ x^- > 0$ are known as the
cosmological regions, where the metric is
\begin{align}
ds^2 = -dt^2 + t^2 d \psi ^2
\end{align}
with the periodicity $\psi \sim \psi + 2 \pi$. Here we performed
coordinate transformations $x^{\pm} = t e^{\pm \psi}$. One region
begins with infinite large volume at far past $t \sim - \infty$,
and meets a big crunch singularity at $t=0$. Another region starts
from big bang singularity at $t=0$ and ends with infinite large
volume at far future $t \sim \infty$. The other two regions with
$x^+x^- < 0$ are the so called whisker regions, where the metric
is given by
\begin{align}
ds^2 = d r^2 - r^2 d \tau^2
\end{align}
with $x^{\pm} = \pm r e^{\pm \tau}$. Due to the periodicity $\tau
\sim \tau + 2\pi$, there are CTCs in these regions. It can be seen
easily that this space has a structure very similar to the model
discussed in this paper.%
\footnote{See \cite{HE} for the singularity structure in Misner space.}
String theory in this space has been investigated, e.g., in
\cite{Nekrasov,Pioline1,Pioline2,Pioline3,Pioline4}.

\begin{figure}
\centerline{\scalebox{0.6}{\includegraphics{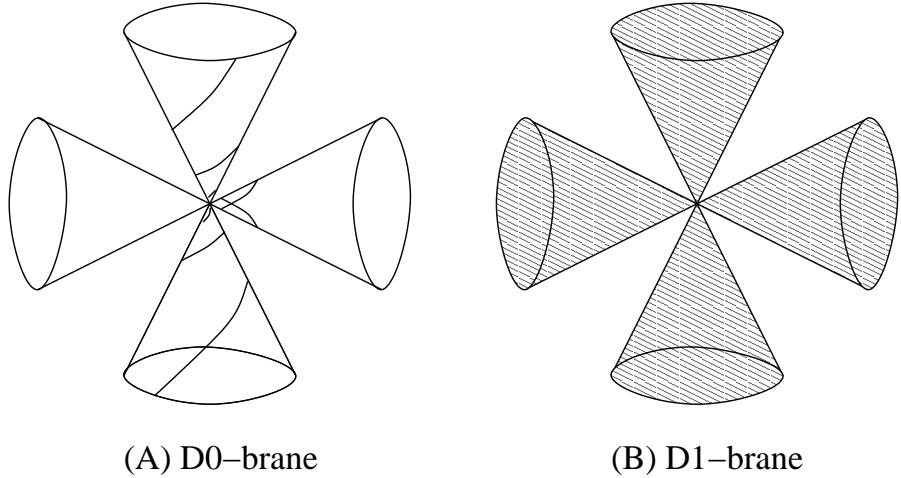}}}
\caption{\it (A) D0-brane propagates from past infinity to future infinity
by passing a whisker region. (B) D1-brane wraps the whole space-time and
it may include gauge flux on it.}
\label{D0D1misner}
\end{figure}
D-branes in Misner space may be obtained by utilizing orbifold
method. In $\br^{1,1}$ we have D0-brane passing from infinite past
to infinite future, and D1-brane with non-trivial gauge flux on
its worldvolume. The trajectory of a particle (D0-brane in our
case) in Misner space is given in  \cite{Pioline4} as follows.
It starts from far past in the cosmological
region $t \sim -\infty$ and approaches spirally into the big crunch
singularity. Then it crosses a whisker region and goes to the
other cosmological region (see figure \ref{D0D1misner}). For
D1-brane, we can construct only brane wraping the whole space-time.
The gauge flux on it is obtained by minimizing the DBI action. In the
cosmological regions we have
\begin{align}
 S = - \tau_1 \int dt d \psi \sqrt{t^2 -F_{t\psi}^2} .
\end{align}
Thus the gauge flux can be computed from the Gauss constraint
\begin{align}
 \frac{\delta {\cal L}}{\delta F_{t\psi}} &=
 \frac{\tau_1 F_{t\psi}}{\sqrt{t^2 - F_{t\psi}^2}} = \Pi,
 &F_{t\psi}^2 &= \frac{t^2}{1 + \frac{\tau_1^2}{ \Pi^2}} .
\end{align}
In the whisker regions, similar analysis leads to the gauge field
\begin{align}
  F_{r\tau}^2 = \frac{r^2}{1 + \frac{\tau_1^2}{\Pi^2}} .
\end{align}
We hope to present a more detailed report on the analysis of the
D-branes including the boundary states for them in near future
\cite{misner-boundary}.

%%%%%%%%%%%%%%%%%%%%%%%%%%%%%%%%%%%%%%%%%%%%%%%%%%%%%%%%%%%%%%%%%%%%%%%%%

\end{document}